\documentclass{JHEP3}
\usepackage{amsmath}
\input epsf
\usepackage{epsfig}
\usepackage{amssymb}
\usepackage{graphics}
\usepackage[active]{srcltx}

\newcommand{\bea}{\begin{eqnarray}}
\newcommand{\eea}{\end{eqnarray}}
\newcommand{\bean}{\begin{eqnarray*}}
\newcommand{\eean}{\end{eqnarray*}}
\newcommand{\nn}{\nonumber \\}

\def\W #1{\widetilde{#1}}
\def\WH #1{\widehat{#1}}

\def\braket#1{\left\langle #1 \right\rangle}

\def\eref#1{(\ref{#1})}

\def\a{{\alpha}}

\def\b{{\beta}}

\def\la{\lambda}
\def\eps{\epsilon}

\def\vev{\braket}

\def\bvev#1{\left[ #1 \right]}
\def\Spaa{\vev}
\def\Spbb{\bvev}

\def\Sl{\sum\limits}
\def\Label#1{\label{#1}%
  \smash{\hbox to0pt{\raise1ex\hbox{\tiny[#1]}\hss}}}

\allowdisplaybreaks
\title{Note on Soft Graviton theorem by KLT Relation}

\author{Yi-Jian Du ${}^{a}$\footnote{On leave from Center for Field Theory
and Particle Physics, Department of Physics, Fudan University, P.R China.}, Bo Feng ${}^{b,c}$, Chih-Hao Fu ${}^{c,d}$, Yihong Wang ${}^e$
 ~~~~~~~~~~~~~~\\${}^a$ Department of Astronomy and Theoretical Physics, Lund University
SE-22362 Lund, Sweden\\
 ${}^b$ Zhejiang Institute of Modern Physics, Zhejiang
University, 38 Zheda Road Hangzhou, 310027 P.R China\\
${}^c$
Center of Mathematical Science, Zhejiang
University, 38 Zheda Road Hangzhou, 310027 P.R China\\
${}^d$ Department of Electrophysics, National Chiao Tung University\\
1001 University Road, Hsinchu, Taiwan, R.O.C.\\
${}^e$ C. N. Yang Institute for Theoretical Physics,
Stony Brook University, Stony Brook, NY 11794.
~~~~~~~\\
 \email{ yijian.du@thep.lu.se; b.feng@cms.zju.edu.cn;\\
 zhihao.fu@cms.zju.edu.cn; yihong.wang@stonybrook.edu \hskip0.5cm} }

\preprint{LU-TP 14-33\\ YITP-SB-14-29}

\date{\today}
\abstract{Recently, new soft graviton theorem proposed by Cachazo and Strominger has inspired a lot of works.
In this note, we use the KLT-formula to investigate the theorem. We have shown 
how the soft  behavior of color ordered Yang-Mills amplitudes can be combined with KLT relation to give the
soft behavior of gravity amplitudes. As a byproduct, we find two nontrivial identities of the KLT momentum kernel  must hold.

}


\begin{document}

\section{Introduction}
Many scattering amplitudes were shown to have an universal soft behavior when the momentum of an external leg tends to zero. The soft limit can be traced back to the work \cite{Low,Weinberg:1964ew}. In recent years, a new soft theorem for gravity amplitudes was studied in
\cite{Strominger:2013jfa,He:2014laa,Kapec:2014opa}. Using Britto-Cachazo-Feng-Witten (BCFW) recursion \cite{Britto:2004ap,Britto:2005fq}
, Cachazo and Strominger have proved the sub- and subsub- leading orders in the soft expansion \cite{Cachazo:2014fwa}, i.e.,\footnote{The leading soft factor $S_{GR}^{(0)}$ is not corrected to all loop orders is shown in \cite{Weinberg:1964ew,Bern:1998sv} while the general subleading behavior of soft gluons and gravitons has also been discussed in \cite{White}.}
\bea {M}_{n+1}(\{\eps \la_s, \W\la_s\},1,...,n) =\left( {1\over \eps^3} S_{GR}^{(0)}+ {1\over \eps^2}
S_{GR}^{(1)}+ {1\over \eps}
S_{GR}^{(2)}\right) { M}_n(1,...,n)+{\cal O}(\eps^0).~~~~\Label{Gra-soft-form}\eea
The leading, subleading and subsubleading orders of soft factors are given by
\bea S_{GR}^{(0)} & = & \sum_{a=1}^n {\varepsilon_{\mu\nu}^{s} p_a^\mu p_a^\nu\over p_s\cdot p_a},~~~S_{GR}^{(1)}= \sum_{a=1}^n {\varepsilon_{\mu\nu}^{s} p_a^\mu (p_{s,\rho} J^{\rho\nu}_a)\over p_s\cdot p_a},~~~
S_{GR}^{(2)}={1\over 2}\sum_{a=1}^n {\varepsilon_{\mu\nu}^{s} (p_{s,\rho} J^{\rho\mu}_a)(p_{s,\sigma} J^{\sigma\nu}_a)\over p_s\cdot p_a},~~~\Label{CS-conj}\eea
where the $\varepsilon_{\mu\nu}^{s}$ is the polarization of the soft graviton, $p_i$ are external momenta and $J^{\mu\nu}$ are angular momenta of external legs. Using the BCFW recursion relation,  the soft limit of  color-ordered tree-level Yang-Mills amplitudes was also studied in \cite{Casali:2014xpa} and the result is given by
\bea A_{n+1}(\{\eps \la_s, \W\la_s\},1,...,n) =\left( {1\over \eps^2} S_{YM}^{(0)}+ {1\over \eps}
S_{YM}^{(1)}\right) A_n(1,...,n),~~~~\Label{YM-soft-form}\eea
where the leading and subleading soft factors  are given by
\bea S_{YM}^{(0)}=\sum_{a\sim s} { \varepsilon_s\cdot p_a\over p_s\cdot p_a},~~ S_{YM}^{(1)}= \sum_{a\sim s} {\varepsilon_{s\nu} p_{s\mu} J^{\mu\nu}_a\over p_s\cdot p_a},~~~\Label{YM-conj} \eea
with $\varepsilon_{s\nu}$ denoting the polarization of the soft gluon and $a\sim s$ meaning partial $a$ is next to soft particle $s$.
Many related studies have been achieved including the soft limits from Poincare symmetry and gauge invariance \cite{Broedel:2014fsa,Bern:2014vva}, Feynman diagram approach \cite{White:2014qia}, conformal symmetry approach to the soft limits in Yang-Mills theory \cite{Larkoski:2014hta}, the soft limit in arbitrary dimension \cite{Schwab:2014xua,Afkhami-Jeddi:2014fia,Zlotnikov:2014sva,Kalousios:2014uva}, loop correction of the soft limit \cite{Bern:2014oka,He:2014bga,Cachazo:2014dia, Bianchi:2014gla}, string-theory approach to the soft limit \cite{Schwab:2014fia,Bianchi:2014gla} and ambitwistor string approach \cite{Adamo:2014yya, Geyer:2014lca}.

In physics, it is very fruitful to study same thing from various angles because it will deepen our understanding and reveal many hidden relations. Now on-shell graviton scattering amplitudes can be
calculated using many different ways, such as BCFW recursion relation, the double-copy formula \cite{Bern:2008qj}, CHY formula \cite{Cachazo:2013hca, Cachazo:2013iea} and KLT formula \cite{KLT} (and many more). Since the BCFW recursion relation and CHY formula  have been successfully used in the study,
in this note we will try to use the KLT formula to investigate the new soft graviton theorem.

Gravity amplitudes at tree level satisfy the famous Kawai-Lewellen-Tye (KLT) relation \cite{KLT}, with which, one can express  the {\bf stripped} tree-level gravity amplitudes $M_n$ (i.e., the momentum conservation  $\delta^4(\sum p_i)$ has been moved away) in terms of products of tree-level color-ordered {\bf stripped} Yang-Mills amplitudes $A_n$ and $\W A_n$
\bea
{M}_{n}(1,2,\dots,n)=\Sl_{\sigma,\rho} A_{n}(\sigma){\cal S[\sigma|\rho]}\W A_{n}(\rho), ~~~\Label{KLT}
\eea
where ${\cal S}[\sigma|\rho]$ is called {\bf momentum kernel}, which is a function of kinematic factors $s_{ij}=2p_i\cdot p_j$ and depends on  the permutations $\sigma$ and $\rho$\footnote{In fact, the momentum kernel can be treated as the metric on the space of $(n-3)!$ BCJ basis. }. KLT relation was firstly proposed in string theory \cite{KLT} and then was proved in field theory  \cite{BjerrumBohr:2010ta,BjerrumBohr:2010yc} using BCFW recursion. One important feature should be emphasized is that KLT is relation between stripped amplitudes without imposing momentum conservation delta function.

Since KLT relation \eqref{KLT} connects gravity amplitudes to Yang-Mills amplitudes, it is natural to expect that the soft limit of gravity amplitudes can be derived from that of Yang-Mills amplitudes  via KLT relation.
In this work, we investigate this connection and its consequences. Although the KLT relation holds to general dimension, for simplicity we will focus on the pure 4D.
 We will show how the leading  and sub-leading soft factors of gravity amplitudes  can be reproduced by the leading and sub-leading soft factors of Yang-Mills amplitudes as it should be. However, to reach such now well established fact, some nontrivial relations among changing matrix of $(n-3)!$ BCJ-basis and momentum kernel ${\cal S}[\rho|\sigma]$ must be true. These nontrivial hidden identities are one of our main results.


The structure of this paper is following. In section 2, we provide a brief review of KLT relation. In section 3, we recall the soft limit for stripped amplitudes of gravity and Yang-Mills theory.
 In section 4, using results in section 3, we present the frame of the proof of the soft graviton soft theorem via KLT relation. In section 5, two examples have been given to demonstrate the frame in section 4. In section 6, we summarize our work with some future directions. In appendix A, we present another more complicated example.

\section{A review of KLT relation}
In this section, we provide a brief review of various formulations of KLT relation for gravity amplitudes (for more details, please refer \cite{BjerrumBohr:2010ta,BjerrumBohr:2010yc}).
The most general formula  \cite{Bern:1998sv} is given as
\begin{align}
{M}_n(1,2,\dots,n) = & (-1)^{n+1}\!\!\! \sum_{\sigma\in S_{n-3}}\sum_{\a\in
S_{j-1}}\sum_{\b\in S_{n-2-j}}
A_n(1,\sigma_{2,j},\sigma_{j+1,n-2},
n-1,n) {\cal S}[
\a_{\sigma(2),\sigma(j)}|\sigma_{2,j}]_{p_1}\nn &
\times {\cal  S}[\sigma_{j+1,n-2}
|\b_{\sigma(j+1),\sigma(n-2)}]_{p_{n-1}} \W
A_n(\a_{\sigma(2),\sigma(j)},1,n-1,
\b_{\sigma(j+1),\sigma(n-2)},n)\,,\Label{KLT-bern}
\end{align}
where $A$ and $\W A$ are two copies of color-ordered Yang-Mills amplitudes and the momentum kernel \cite{BjerrumBohr:2010ta,BjerrumBohr:2010yc,BjerrumBohr:2010hn} is defined as
\bea {\cal S}[i_1,i_2,...,i_k|j_1,j_2,...,j_k]_{p_1} & = & \prod_{t=1}^k
(s_{i_t 1}+\sum_{q>t}^k \theta(i_t,i_q) s_{i_t
i_q})~~~\Label{S-def}\eea
where $p_1$ is the pivot and $\theta(i_t, i_q)$ is zero when  pair $(i_t,i_q)$ has same
ordering at both set ${\cal I}=\{i_1,i_2,...,i_k\},{\cal J}=\{j_1,j_2,...,j_k\}$, otherwise it is one\footnote{
The function ${\cal S}$ is nothing, but the $f$-function defined in
\cite{Bern:1998sv} with more symmetric and improved expression}. In this definition, the
set ${\cal J}=\{j_1,j_2,...,j_k\}$ is the reference ordering set,
i.e., this set provides the standard ordering. The set ${\cal
I}=\{i_1,i_2,...,i_k\}$ is the dynamical set which determines the
dynamical factor by comparing with set ${\cal J}$.
A few examples are the 
following:
\bean {\cal S}[2,3,4|2,4,3]_{p_1} & = & s_{21} (s_{31}+s_{34}) s_{41},~~~
{\cal S}[2,3,4|4,3,2]_{p_1}=(s_{21}+s_{23}+s_{24})(s_{31}+s_{34}) s_{41}.
\eean
Although it is not so obvious, the momentum kernel, in fact, contains all BCJ-relations by following identities
\bea 0=\sum_{\a\in S_{n-2}}{\cal S}[\a(i_2,...,i_{n-1})|j_2,j_3,...,j_{n-2}]
A_n(n,\a(i_2,...,i_{n-1}),1)~,\forall j\in S_{n-2}~~~\Label{I-com}\eea
Using \eref{I-com} we can derive
 following relation
\begin{align}
&\sum_{\alpha,\beta} \mathcal{S}[\alpha_{i_2,i_j}|i_2,\ldots,i_j]_{p_1}
{\mathcal{S}}[i_{j+1},\ldots,i_{n-2}|\beta_{i_{j+1},i_{n-2}}]_{p_{n-1}}
\widetilde{A}_n(\alpha_{i_2,i_j},1,n-1,\beta_{i_{j+1},i_{n-2}},n) \nonumber \\
&=\sum_{\alpha',\beta'} \mathcal{S}[\alpha'_{i_2,i_{j-1}}|i_2,\ldots,i_{j-1}]_{p_1}
{\mathcal{S}}[i_j,i_{j+1},\ldots,i_{n-2}|\beta'_{i_j,i_{n-2}}]
\widetilde{A}_n(\alpha'_{i_2,i_{j-1}},1,n-1,\beta'_{i_j,i_{n-2}},n)\,,
\Label{jrel}
\end{align}
Thus we can shift
$j$ in \eref{KLT-bern} all the way to make the left- or right-hand part empty,
\textit{i.e.} we can choose $j=1$ or $j=n-2$. These special cases corresponds to the manifest $S_{n-3}$-symmetric form \eqref{KLT-gra} and its dual form \eqref{dual-KLT-gra},
which are given by
\bea &&{M}_n(1,...,n)=
(-)^{n+1}\sum_{\sigma,\W\sigma\in S_{n-3}} A_n(1,\sigma_{2,n-2},
n-1,n) {\cal S}[ \W\sigma_{2,n-2}|\sigma_{2,n-2}]_{p_1} \W
A_n(n-1,n,\W\sigma_{2,n-2},1).~~\Label{KLT-gra}\eea
and
\bea
&&{M}_n(1,\dots,n)=(-1)^{n+1}\sum_{\sigma,\widetilde{\sigma} \in S_{n-3}}
A_n(1,\sigma_{2,n-2},n-1,n)
{\mathcal{S}}[\sigma_{2,n-2}|\widetilde{\sigma}_{2,n-2}]_{p_{n-1}}
\widetilde{A}_n(1,n-1,\widetilde{\sigma}_{2,n-2},n).
~~\Label{dual-KLT-gra}
\eea
%

\section{Review of soft limits of gravity and Yang-Mills theory}
In this section, we review the soft behavior of gravity and Yang-Mills theory given in \cite{Casali:2014xpa, Cachazo:2014fwa}. Since in KLT formula, amplitudes used are these {\bf stripped} amplitudes, thus we will focus on the soft behaviors of these amplitudes.


We focus on the four dimensional case, thus we can use  spinor variables. Under these variables, soft factors in \eref{CS-conj} and \eref{YM-conj}
are given by \cite{Cachazo:2014fwa} for gravity theory\footnote{It is worth to emphasize that here we have used the QCD convention, i.e., $2p\cdot q=\Spaa{p|q}\Spbb{q|p}$.}
\bea S_{GR}^{(0)}& = & -\sum_{i=1}^n {\Spbb{s|i}\Spaa{x|i}\Spaa{y|i}\over \Spaa{s|i}\Spaa{x|s}\Spaa{y|s}},~~~~~ S_{GR}^{(1)}=-{1\over 2}\sum_{i=1}^n {\Spbb{s|i}\over \Spaa{s|i}}\left( {\Spaa{x|i}\over \Spaa{x|s}}+{\Spaa{y|i}\over \Spaa{y|s}}\right) \W\la_s^{\dot{\a}}{\partial \over \partial\W\la_i^{\dot{\a}}}\nn
 S_{GR}^{(2)}& = & -{1\over 2}\sum_{i=1}^n {\Spbb{s|i}\over \Spaa{s|i}}\W\la_s^{\dot{\a}}\W\la_s^{\dot{\b}}{\partial^2 \over \partial\W\la_i^{\dot{\a}}\W\la_i^{\dot{\b}}},~~~\Label{CS-conj-4D} \eea
where $x,y$ are two auxiliary spinors used to define the helicity of soft graviton
\bea \epsilon^{+2}= \left({\la_x \W\la_k\over \Spaa{x|k}}\right) \left({\la_y \W\la_k\over \Spaa{y|k}}\right)+\{x\leftrightarrow y\}~,\eea
 and  by \cite{Casali:2014xpa} for Yang-Mills theory\footnote{We have assumed the color ordering is
 $(1,...,n, s)$. }
\bea S_{YM}^{(0)}(n,s,1,...) & = & {\Spaa{n|1}\over \Spaa{n|s}\Spaa{s|1}},~~~~
S_{YM}^{(1)}(n,s,1,...)
 =  {1\over
\Spaa{s|1}}\W\la_s
{\partial \over \partial \W\la_1}+{1\over \Spaa{n|s}}\W\la_s {\partial \over
\partial \W\la_n}.~~~~\Label{My-soft-gluon-new} \eea
To reach these expressions, we have used the fact that  in 4D, angular momentum can be written as spinor form
\bea J_{\mu\nu}\to J_{\a \b} \eps_{\dot{\a}\dot{\b}}+\W J_{\dot{\a}\dot{\b}} \eps_{\a\b},~~~
 J_{\a \b}=\la_\a{\partial \over \partial \la^\b}+\la_\b{\partial \over \partial \la^\a},
 ~~~ \W J_{\dot{\a }\dot {\b}}=\W\la_{\dot \a}{\partial \over \partial \W\la^{\dot\b}}+\W\la_{\dot \b}{\partial \over \partial\W \la^{\dot \a}}.~~~\Label{Angular-4D} \eea
We will explain the meaning of differential operators for stripped amplitudes shortly.

For stripped amplitudes, we must impose momentum conservation from beginning. This can be done as given in \cite{Cachazo:2014fwa}.
Under the {\bf holomorphic soft limit} which is defined as
\bea \la_s\to \eps \la_s,~~~\W\la_s\to \W\la_s~~~~\Label{kin-soft-2}  \eea
momentum conservation  $\sum_{i=1}^n k_i+\eps k_s=0$ can be used to
 solve two arbitrarily chosen anti-spinors $\W\la_i, \W\la_j$ as
\bea \W\la_i=-\sum_{k\neq i,j} {\Spaa{j|k}\over \Spaa{j|i}}\W\la_k-\eps{\Spaa{j|s}\over \Spaa{j|i}}\W\la_s ,~~~\W\la_j=-\sum_{k\neq i,j} {\Spaa{i|k}\over \Spaa{i|j}}\W\la_k-\eps{\Spaa{i|s}\over \Spaa{i|j}}\W\la_s
~~~~\Label{kin-soft-3}\eea
In other words, for stripped amplitudes, now
the independent variables are $\la_i$( $i=1,...,n$), $\la_s$, $\W\la_s$ and
$\W\la_k$ ($k=1,...,n$ and $k\neq i,j$). With the fixed choice of pair $(i,j)$, when we use the BCFW recursion relation to discuss the soft behavior as was done in \cite{He:2014bga}, for example, for an $(n+1)$-point color-ordered Yang-Mills amplitude $A(\{\eps \la_s,\W\la_s\}, \{\la_1,\W\la_1\},...,\{\la_n,\W\la_n\})$
with  $h_s=+1$, we will receive contributions to the singular part from the two-particle channel
\bea A_{n+1}\left(\{\eps \la_s, \W\la_s\}^+,1,...,n\right)|_{div}
& = &  A_3\left(\WH s^+, 1^{h_1}, -\WH P_{1s}^{-h_i}\right) {1\over P_{1s}^2}
A_n\left(\WH P_{1s}^{h_i},...,\WH n\right)\mid_{div}~~~\Label{soft-div-11} \eea
under  the $(s,n)$-shift
\bea \eps\la_s(z)=\eps\la_s+ z\la_n,~~~~~\W\la_n(z)=\W\la_n-z\W\la_s~.~~~\Label{soft-def}\eea
It is easy to calculate  the divergent part and we find
\bea
 {- \Spaa{n|1}\over \eps^2\Spaa{n|s}\Spaa{s|1}}
A_{n} \left(\{\la_1, \W\la_1+ \eps{\Spaa{n|s}\over \Spaa{n|1}}\W\la_s\}^{h_1},...,\{\la_i, \W\la_i(\eps)\},...\{\la_j,\W\la_j(\eps)\},...,\{\la_n, \W\la_n+\eps{\Spaa{1|s}\over \Spaa{1|n}}\W\la_s \}\right)\eea
where \eref{kin-soft-3} must  be used. A compact way to rewrite above expression is to assume $\W\la_i, \W\la_j$ to be independent first, so we have
\bea
 {- \Spaa{n|1}\over \eps^2\Spaa{n|s}\Spaa{s|1}}\left\{ {{\bf e}^{ \eps{\Spaa{n|s}\over \Spaa{n|1}}\W\la_s{\partial \over \partial \W\la_1}-\eps{\Spaa{j|s}\over \Spaa{j|i}}\W\la_s {\partial \over \partial \W\la_i} -\eps{\Spaa{i|s}\over \Spaa{i|j}}\W\la_s {\partial \over \partial \W\la_j}+{\eps\Spaa{1|s}\over \Spaa{1|n}}\W\la_s {\partial \over \partial \W\la_n}  }} A_{n} \left(\{\la_1, \W\la_1\},\dots,\{\la_{j},\W\la_{j}\},\dots,\{\la_n, \W\la_n \}\right)\right\}~~~\Label{soft-div-13-4}\eea
Only after the action of ${\partial \over \partial \W\la_i}$ and ${\partial \over \partial \W\la_j}$, we can replace $\W\la_i, \W\la_j$ by \eref{kin-soft-3} with $\eps=0$. However, if we insist to use \eref{kin-soft-3} from beginning,  $\W\la_i, \W\la_j$ will depend on $\W\la_1, \W\la_n$
thus  the total
derivative of ${d\over d\W\la_1}$ and ${d\over d\W\la_n}$ must be written as
\bea {d\over d\W\la_1} & = &  {\partial\over \partial\W\la_1}+{\partial\over \partial\W\la_i} \left(
-{\Spaa{j|1}\over \Spaa{j|i}}\right)+{\partial\over \partial\W\la_j} \left(
-{\Spaa{i|1}\over \Spaa{i|j}}\right)\nn
{d\over d\W\la_n} & = &  {\partial\over \partial\W\la_n}+{\partial\over \partial\W\la_i} \left(
-{\Spaa{j|n}\over \Spaa{j|i}}\right)+{\partial\over \partial\W\la_j} \left(
-{\Spaa{i|n}\over \Spaa{i|j}}\right)~.~~~\Label{total-wla1n}\eea
Using above formula, it is easy to check that
\bea{\Spaa{n|s}\over
\Spaa{n|1}}\W\la_s
{d \over d \W\la_1}+{\Spaa{1|s}\over \Spaa{1|n}}\W\la_s {d \over
d \W\la_n}
& = & {\Spaa{n|s}\over
\Spaa{n|1}}\W\la_s
{\partial \over \partial \W\la_1}+{\Spaa{1|s}\over \Spaa{1|n}}\W\la_s {\partial \over
\partial \W\la_n}- {\Spaa{j|s}\over \Spaa{j|i}}\W\la_s{\partial\over \partial\W\la_i}-{\Spaa{i|s}\over \Spaa{i|j}}\W\la_s{\partial\over \partial\W\la_j}, \eea
thus \eref{soft-div-13-4} becomes
\bea
 {- \Spaa{n|1}\over \eps^2\Spaa{n|s}\Spaa{s|1}}\left\{ {\bf e}^{ {\Spaa{n|s}\over
\Spaa{n|1}}\W\la_s
{d \over d \W\la_1}+{\Spaa{1|s}\over \Spaa{1|n}}\W\la_s {d \over
d \W\la_n}} A_{n} \left(\{\la_1, \W\la_1\},\dots,\{\la_{j},\W\la_{j}\},\dots,\{\la_n, \W\la_n \}\right)\right\}~~~\Label{soft-div-13-5}\eea
Having this new understanding, the meaning of soft factors in \eref{CS-conj-4D} and \eref{My-soft-gluon-new} becomes clear: {\sl while there are no variables $\W\la_i, \W\la_j$ anymore in stripped amplitudes, all partial derivatives should be considered as a kind of "total derivative" in the sense of \eref{total-wla1n}.}

\section{ KLT relation approach to the soft behavior of gravity amplitude}

Having above preparations, now we study the soft behavior of stripped gravity amplitudes using the soft behavior of stripped Yang-Mills amplitudes as input  through KLT relation. The total symmetry among
the $n$-particles of gravity amplitudes allows us to choose any leg to be soft leg. We take $p_1$ to
be soft and solve $n-1, n$ as
\bea \W\la_{n-1}=-\sum_{k=2}^{n-2} {\Spaa{n|k}\over
\Spaa{n|n-1}}\W\la_k-\eps{\Spaa{n|1}\over \Spaa{n|n-1}}\W\la_1
,~~~\W\la_n=-\sum_{k=2}^{n-2} {\Spaa{n-1|k}\over
\Spaa{n-1|n}}\W\la_k-\eps{\Spaa{n-1|1}\over \Spaa{n-1|n}}\W\la_1.
~~~~\Label{Solve-nn-1}\eea

{\bf The choice of  KLT formula:} In section 2, we have reviewed  various formulations of KLT relation.
To make the discussion simpler, we should start with proper choice of KLT formula. Since the leading contribution from two gluon amplitudes is the order ${1\over \eps^2}\times {1\over \eps^2}$ while the leading contribution of graviton amplitude is ${1\over \eps^3}$, we are better to have manifest $\eps$-factor from kernel part. Furthermore, since we have solved $\W\la_{n-1},\W\la_n$ in
\eref{Solve-nn-1}, it is more convenient to have formula as less
related to $p_{n-1}, p_n$ as possible. Taking these things into consideration, we use the general formula given
by \eref{KLT-bern} with $j=2$
\bea M_n &= & (-1)^{n+1} \sum_{t=2}^{n-2}\sum_{\sigma,\b\in S_{n-4}}
A_n(1,t,\sigma,n-1,n) {\cal S}[t|t]_{p_1}{\cal
S}[\sigma|\b]_{p_{n-1}} \W
A_n(t,1,n-1,\b,n)~~~~~\Label{KLT-Bern-j=2} \eea
In this form, ${\cal S}[t|t]_{p_1} \to \eps s_{1t}$, while
 the expansion of the other
kernel ${\cal
S}[\sigma|\b]_{p_{n-1}}$ can be written as\footnote{From the definition of kernel, the $\eps$-expansion
should be given by ${\bf e}^{-\eps {\Spaa{n|1} \over \Spaa{n|n-1}} {\W
\la_1|{\partial \over \partial \W \la_{n-1}}}}{\cal
S}[\sigma|\b]_{p_{n-1}}$. However, noticing  that
\bean & & \W\la_1 {d\over d\W\la_t} {\cal  S}[\sigma|\b]_{p_{n-1}}
= \W\la_1 \left( -{\Spaa{n|t}\over \Spaa{n|n-1}}\right){\partial\over \partial \W\la_{n-1}} {\cal  S}[\sigma|\b]_{p_{n-1}}\eean
where we have used the fact that $\W\la_s {d\over d\W\la_t} {\cal  S}[\sigma|\b]_{p_{n-1}}$
does not contain momentum $p_t$, we obtain \eref{Soft-kernel-2}. }
\bea & &  {\bf e}^{+\eps  {\Spaa{n|1}\over \Spaa{n|t}} \W\la_1 {d\over d\W\la_t}  }{\cal
S}[\sigma|\b]_{p_{n-1}}. \Label{Soft-kernel-2}\eea
For convenience, we use \eqref{My-soft-gluon-new} to write down the singular soft limit of two stripped amplitudes
in \eref{KLT-Bern-j=2} as
\bea   A_n^{(n-1,n)}(1,t,\sigma,n-1,n)
& \to & {1\over \eps^2}
{\Spaa{n|t}\over \Spaa{n|1}\Spaa{1|t}}
A_{n-1}^{(n-1,n)}(t,\sigma,n-1,n)\nn
&&+ {1\over \eps} {\Spaa{n|t}\over \Spaa{n|1}\Spaa{1|t}} \left( {\Spaa{n|1}\over
\Spaa{n|t}}\W\la_1
{d \over d \W\la_t}+{\Spaa{t|1}\over \Spaa{t|n}}\W\la_1 {d \over
d \W\la_n}\right)A_{n-1}(t,\sigma,n-1,n),~~\Label{My-A-soft}
\eea
\bea \W A_n^{(n-1,n)}(t,1,n-1,\b,n)& \to & {1\over \eps^2}
{\Spaa{t|n-1}\over \Spaa{t|1}\Spaa{1|n-1}}\W
A_n(t,n-1,\b,n)\nn
&&+ {1\over \eps} {\Spaa{t|n-1}\over
\Spaa{t|1}\Spaa{1|n-1}}\left(
{\Spaa{t|1}\over \Spaa{t|n-1}}\W\la_1{d\over d
\W\la_{n-1}} +{\Spaa{n-1|1}\over \Spaa{n-1|t}}\W\la_1 {d
\over d \W\la_t}\right)\W
A_n(t,n-1,\b,n).\Label{My-At-soft}\eea

In the remainder of this section, we discuss the soft behavior of gravity amplitudes by KLT relations order by order.
\subsection{The leading order part}
Substituting the leading part of color-ordered Yang-Mills amplitudes $A$, $\W A$ (given by ${1\over \eps^2}$ terms of \eqref{My-A-soft}, \eqref{My-At-soft}) as well as the leading part of momentum kernel $\cal S$ (given by the $\eps$ term of ${\cal S}[t|t]_{p_1}{\cal
S}[\sigma|\b]_{p_{n-1}}$) into the KLT expression \eref{KLT-Bern-j=2}, we get the leading part of gravity amplitude under soft limit
\bea M_n &= & (-1)^{n+1} \sum_{t=2}^{n-2}\sum_{\sigma,\b\in S_{n-4}}
{1\over \eps^2}
{\Spaa{n|t}\over \Spaa{n|1}\Spaa{1|t}}
A_{n-1}^{(n-1,n)}(t,\sigma,n-1,n)\eps s_{1t} {\cal
S}[\sigma|\b]_{p_{n-1}^{\eps\to 0}}\nn & & {1\over \eps^2}
{\Spaa{t|n-1}\over \Spaa{t|1}\Spaa{1|n-1}}\W
A_n^{(n-1,n)}(t,n-1,\b,n)\nn
& = & {1\over \eps^3} (-1)^{n+1} \sum_{t=2}^{n-2}
{\Spbb{t|1}\over \Spaa{t|1}}{\Spaa{n|t}\over \Spaa{n|1}}{\Spaa{n-1|t}\over \Spaa{n-1|1}}
\sum_{\sigma,\b\in S_{n-4}} A_{n-1}^{(n-1,n)}(t,\sigma,n-1,n) {\cal
S}[\sigma|\b]_{p_{n-1}^{\eps\to 0}}\W
A_n^{(n-1,n)}(t,n-1,\b,n)\nn
& = & {1\over \eps^3} (-) \sum_{t=2}^{n-2}
{\Spbb{t|1}\over \Spaa{t|1}}{\Spaa{n|t}\over \Spaa{n|1}}{\Spaa{n-1|t}\over \Spaa{n-1|1}}
M_{n-1}(2,...,n)\nn
& = &{1\over \eps^3} S^{(0)}_{GR}
M_{n-1}(2,...,n) \eea
where, on the third line, we have used the $S_{n-3}$-symmetric KLT relation
\eref{dual-KLT-gra}  for $(n-1)$-point amplitudes. The soft factor of gravity is nothing but the $S^{(0)}_{GR}$ defined in \eqref{CS-conj-4D} with $x=n$ and $y=n-1$.

\subsection{The subleading order part}
Now let us study the subleading order of stripped gravity amplitudes under the soft limit.
We will do it in three steps. In the first step, we act the
  $S_{GR}^{(1)}$ defined in \eref{CS-conj-4D}  on the KLT expressions  \eref{dual-KLT-gra}  of $(n-1)$-point gravity amplitudes directly.
  In the second step, we collect
 contributions of the subleading part from color ordered Yang-Mills amplitudes and momentum kernel
 in \eref{KLT-Bern-j=2}. Finally, we compare the two expressions from first two steps to prove (check) the subleading order soft factor $S_{GR}^{(1)}$  of gravity amplitude.

\subsubsection{The sub-leading part from direct acting of $S_{GR}^{(1)}$}
We use the subleading soft factor given by \eref{CS-conj-4D}
  \bea S_{GR,(n-1)n}^{(1)}& = & - \sum_{i=2}^{n} {\Spbb{1|i}\over
\Spaa{1|i}}{\Spaa{n-1|i}\over\Spaa{n-1|1}}
   \W\la_1 {d \over d \W\la_i}= - \sum_{i=2}^{n-2} {\Spbb{1|i}\over
\Spaa{1|i}}{\Spaa{n-1|i}\over\Spaa{n-1|1}}
   \W\la_1 {d \over d \W\la_i}~~~~\Label{MyKLT-direct-S11-1}  \eea
   where we have taken the gauge choice $x=y=n-1$, thus ${d \over d \W\la_{n-1}}=
   {d \over d \W\la_n}=0$. When acting it with the  form \eqref{MyKLT-direct-S11-1} on $M_{n-1}$,
for each $i$, we take different representation of $M_{n-1}$\footnote{It is worth to notice that although
as a whole, we have the freedom to chose $x,y$ for $S_{GR,(n-1)n}^{(1)}$, when we act it for different $i$ and different part $A,\W A$ in \eref{Direct-2}, we need to stick to a particular  gauge choice. }, i.e.,
\bea  & & S_{GR,(n-1)n}^{(1)} M_{n-1}(2,...,n)=- \sum_{i=2}^{n-2} {\Spbb{1|i}\over
\Spaa{1|i}}{\Spaa{n-1|i}\over\Spaa{n-1|1}}
   \W\la_1 {d \over d \W\la_i}  M_{n-1}(2,...,n)\nn
&= &  - \sum_{i=2}^{n-2} {\Spbb{1|i}\over
\Spaa{1|i}}{\Spaa{n-1|i}\over\Spaa{n-1|1}}
   \W\la_1 {d \over d \W\la_i}\left[(-1)^{n} \sum_{\sigma, \b\in S_{n-4}}A_{n-1}(i,\sigma,n-1,n) {\cal
S}[\sigma|\b]_{p_{n-1}} \W A_{n-1}(i,n-1,\b,n)\right]\nn
 &= &  (-)^{n+1} \sum_{i=2}^{n-2} \sum_{\sigma, \b\in S_{n-4}} {\Spbb{1|i}\over
\Spaa{1|i}}{\Spaa{n-1|i}\over\Spaa{n-1|1}}
    A_{n-1}(i,\sigma,n-1,n) \left(\W\la_1 {d \over d \W\la_i} {\cal
S}[\sigma|\b]_{p_{n-1}}\right) \W A_{n-1}(i,n-1,\b,n)\nn
& + & (-)^{n+1} \sum_{i=2}^{n-2} \sum_{\sigma, \b\in S_{n-4}} {\Spbb{1|i}\over
\Spaa{1|i}}{\Spaa{n-1|i}\over\Spaa{n-1|1}}
   \left(\W\la_1 {d \over d \W\la_i} A_{n-1}(i,\sigma,n-1,n) \right) {\cal
S}[\sigma|\b]_{p_{n-1}} \W A_{n-1}(i,n-1,\b,n)\nn
& + & (-)^{n+1} \sum_{i=2}^{n-2} \sum_{\sigma, \b\in S_{n-4}} {\Spbb{1|i}\over
\Spaa{1|i}}{\Spaa{n-1|i}\over\Spaa{n-1|1}}
    A_{n-1}(i,\sigma,n-1,n) {\cal
S}[\sigma|\b]_{p_{n-1}} \left(\W\la_1 {d \over d \W\la_i}\W A_{n-1}(i,n-1,\b,n)\right).~~~\Label{Direct-2}\eea

\subsubsection{The sub-leading order part from KLT relation}
Now we collect the contributions of the subleading part from the KLT relation \eqref{KLT-Bern-j=2}.
There are three contributions at this order. The first term is to
take kernel to second order of $\eps$, while $A,\W A$ are the first
order (see \eref{My-A-soft} and \eref{My-At-soft}). This part is given by
\bea T_1 & = & (-1)^{n+1} \sum_{t=2}^{n-2}\sum_{\sigma,\b\in S_{n-4}}
{\Spbb{1|t}\over \Spaa{1|t}} {\Spaa{n|t}\Spaa{n-1|t}\over
\Spaa{n|1}\Spaa{n-1|1}}A_{n-1}(t,\sigma,n-1,n) \left({\Spaa{n|1}\over \Spaa{n|t}} \W\la_1 {d\over d\W\la_t}{\cal
S}[\sigma|\b]_{p_{n-1}}\right)   \W A_{n-1}(t,n-1,\b,n) \nn
& = & (-1)^{n+1} \sum_{t=2}^{n-2}\sum_{\sigma,\b\in S_{n-4}}
{\Spbb{1|t}\over \Spaa{1|t}} {\Spaa{n-1|t}\over
\Spaa{n-1|1}}A_{n-1}(t,\sigma,n-1,n) \left( \W\la_1 {d\over d\W\la_t}{\cal
S}[\sigma|\b]_{p_{n-1}}\right)   \W A_{n-1}(t,n-1,\b,n)\eea

For the second term, we keep the leading order of kernel and $\W A$
while taking the subleading order of $A$, thus we have
\bea T_2 & = & (-1)^{n+1} \sum_{t=2}^{n-2}\sum_{\sigma,\b\in S_{n-4}}
{\Spbb{1|t}\over \Spaa{1|t}} {\Spaa{n|t}\Spaa{n-1|t}\over
\Spaa{n|1}\Spaa{n-1|1}}\nn
&&\times \left[\left( {\Spaa{n|1}\over
\Spaa{n|t}}\W\la_1
{d \over d \W\la_t}+{\Spaa{t|1}\over \Spaa{t|n}}\W\la_1 {d \over
d \W\la_n} \right)
A_{n-1}(t,\sigma,n-1,n)\right] {\cal
S}[\sigma|\b]_{p_{n-1}}  \W A_n(t,n-1,\b,n)\nn
& = & (-1)^{n+1} \sum_{t=2}^{n-2}\sum_{\sigma,\b\in S_{n-4}}
{\Spbb{1|t}\over \Spaa{1|t}} {\Spaa{n-1|t}\over \Spaa{n-1|1}} \left(\W\la_1
{d \over d \W\la_t}
A_{n-1}(t,\sigma,n-1,n)\right)  {\cal
S}[\sigma|\b]_{p_{n-1}}  \left( \W A_n(t,n-1,\b,n)\right),\nn\eea
where we have used the fact that ${d \over
d \W\la_n} A_{n-1}(t,\sigma,n-1,n)=0$

For the third term, we keep the leading order of kernel and $ A$
while take the subleading order of $\W A$, thus we have
\bea T_3 & = & (-1)^{n+1} \sum_{t=2}^{n-2}\sum_{\sigma,\b\in S_{n-4}}
{\Spbb{1|t}\over \Spaa{1|t}} {\Spaa{n|t}\Spaa{n-1|t}\over
\Spaa{n|1}\Spaa{n-1|1}}A_{n-1}(t,\sigma,n-1,n) {\cal
S}[\sigma|\b]_{p_{n-1}} \nn
& &  \left(\left\{
{\Spaa{t|1}\over \Spaa{t|n-1}}\W\la_1{d\over d
\W\la_{n-1}} +{\Spaa{n-1|1}\over \Spaa{n-1|t}}\W\la_1 {d
\over d \W\la_t}\right\}\W
A_n^{(n-1,n)}(t,n-1,\b,n)\right)\nn
& = & (-1)^{n+1} \sum_{t=2}^{n-2}\sum_{\sigma,\b\in S_{n-4}}
{\Spbb{1|t}\over \Spaa{1|t}} {\Spaa{n|t}\over
\Spaa{n|1}}A_{n-1}(t,\sigma,n-1,n) {\cal
S}[\sigma|\b]_{p_{n-1}} \left(\W\la_1 {d
\over d \W\la_t}\W
A_n^{(n-1,n)}(t,n-1,\b,n)\right)~~~~~\eea
where again we have used the fact $\W\la_1{d\over d
\W\la_{n-1}}\W A_n^{(n-1,n)}(t,n-1,\b,n)=0$.

\subsubsection{Comparing sub-leading parts}

Now we compare \eref{Direct-2} with $T_1, T_2, T_3$. It is easy to
see when we identify $i=t$, we have
\bea \Delta &= & S_{GR,(n-1)n}^{(1)} M_{n-1}(2,...,n)-T_1-T_2-T_3 \nn
& = & (-1)^{n+1} \sum_{t=2}^{n-2}\sum_{\sigma,\b\in S_{n-4}}
{\Spbb{1|t}\over \Spaa{1|t}} \left({\Spaa{n-1|t}\over \Spaa{n-1|1}}- {\Spaa{n|t}\over
\Spaa{n|1}}\right)A_{n-1}(t,\sigma,n-1,n) {\cal
S}[\sigma|\b]_{p_{n-1}} \left(\W\la_1 {d
\over d \W\la_t}\W
A_n^{(n-1,n)}(t,n-1,\b,n)\right)\nn
&= & (-1)^{n+1} \sum_{t=2}^{n-2}\sum_{\sigma,\b\in S_{n-4}}
{\Spbb{1|t} \Spaa{n|n-1}\over \Spaa{n-1|1}
\Spaa{n|1}}A_{n-1}(t,\sigma,n-1,n) {\cal
S}[\sigma|\b]_{p_{n-1}} \left(\W\la_1 {d
\over d \W\la_t}\W
A_n^{(n-1,n)}(t,n-1,\b,n)\right)\nn
&= & (-1)^{n+1} \sum_{t=2}^{n-2}\sum_{\sigma,\b\in S_{n-4}}
{\W\la_1^{\dot \a}\W\la_1^{\dot \b}  \Spaa{n|n-1}\over \Spaa{n-1|1}
\Spaa{n|1}}A_{n-1}(t,\sigma,n-1,n) {\cal
S}[\sigma|\b]_{p_{n-1}} \left(\W\la_{t,\dot \a} {d
\over d \W\la_t^{\dot \b}}\W
A_n^{(n-1,n)}(t,n-1,\b,n)\right)\nn
 &= & (-1)^{n+1} \sum_{t=2}^{n-2}\sum_{\sigma,\b\in S_{n-4}}
{\W\la_1^{\dot \a}\W\la_1^{\dot \b}  \Spaa{n|n-1}\over \Spaa{n-1|1}
\Spaa{n|1}}A_{n-1}(t,\sigma,n-1,n) {\cal
S}[\sigma|\b]_{p_{n-1}} \left(J_{t,\dot \a \dot \b}\W
A_n(t,n-1,\b,n)\right)~~~~\Label{Need-to-proof}\eea
It is obviously that to prove (or check) the subleading soft factor $S_{GR,(n-1)n}^{(1)}$, we need
to prove (or check)  $\Delta=0$. Before going to the detail, let us notice that in \eref{Need-to-proof}
 only the anti-spinor part of angular momentum $J_{t,\dot \a \dot \b}$ appears.

 Now we present the idea of proof. In \eqref{Need-to-proof}, for each $t$, we have used different BCJ-basis
 for color ordered partial amplitudes. Thus the first step is to translate various basis into a standard basis. In other words, we should do following transformation
%
\bea A_{n-1}(t,\sigma_{t},n-1,n) & = &  \sum_{\sigma_{ \W t}\in S_{n-4}} A_{n-1}(\W t,\sigma_{\W t},n-1,n)
{\cal D}[\W t,\sigma_{\W t},n-1,n|t,\sigma_{t},n-1,n]\nn
\W A_{n-1}(t,n-1,\b_{t},n)& = &\sum_{\b_{ \W t}\in S_{n-4}} {\cal C}[t,n-1,\b_{ t},n|\W t,n-1,\b_{\W t},n]
\W
A_{n-1}(\W t,n-1,\b_{\W t},n).~~\Label{minimal-basis} \eea
where we have used the $\sigma_t$ to denote the permutations of $n-4$-elements after deleting particles $1,n,n-1,t$.
Inserting above transformation into the extra term \eqref{Need-to-proof}, when we choose e.g., $\W t=2$ in above equations, we obtain
\bea&& (-1)^{n+1}\Delta \nn
& = & \sum_{\sigma_2,\b_2\in S_{n-4}}
{\W\la_1^{\dot \a}\W\la_1^{\dot \b}  \Spaa{n|n-1}\over \Spaa{n-1|1}
\Spaa{n|1}}A_{n-1}(2,\sigma_2,n-1,n) \sum_{\W t=2}^{n-2}\sum_{\sigma_{\W t},\b_{\W t}\in S_{n-4}}
{\cal D}[2,\sigma_2,n-1,n|\W t,\sigma_{\W t},n-1,n] {\cal
S}[\sigma_{\W t}|\b_{\W t}]_{p_{n-1}} \nn & &   J_{\W t,\dot \a \dot \b}\left\{{\cal C}[\W t,n-1,\b_{\W t},n|2,n-1,\b_2,n]\W
A_{n-1}(2,n-1,\b_2,n)\right\}\nn
& = & \sum_{\sigma_2,\b_2\in S_{n-4}}
{\W\la_1^{\dot \a}\W\la_1^{\dot \b}  \Spaa{n|n-1}\over \Spaa{n-1|1}
\Spaa{n|1}}A_{n-1}(2,\sigma_2,n-1,n) \sum_{\W t=2}^{n-2}\sum_{\sigma_{\W t},\b_{\W t}\in S_{n-4}}
{\cal D}[2,\sigma_2,n-1,n|\W t,\sigma_{\W t},n-1,n] {\cal
S}[\sigma_{\W t}|\b_{\W t}]_{p_{n-1}} \nn & &   {\cal C}[\W t,n-1,\b_{\W t},n|2,n-1,\b_2,n]\left\{\W
J_{\W t,\dot \a \dot \b}A_{n-1}(2,n-1,\b_2,n)\right\}\nn
& + &  \sum_{\sigma_2,\b_2\in S_{n-4}}
{\W\la_1^{\dot \a}\W\la_1^{\dot \b}  \Spaa{n|n-1}\over \Spaa{n-1|1}
\Spaa{n|1}}A_{n-1}(2,\sigma_2,n-1,n) \sum_{\W t=2}^{n-2}\sum_{\sigma_{\W t},\b_{\W t}\in S_{n-4}}
{\cal D}[2,\sigma_2,n-1,n|\W t,\sigma_{\W t},n-1,n] {\cal
S}[\sigma_{\W t}|\b_{\W t}]_{p_{n-1}} \nn & &   \left\{J_{\W t,\dot \a \dot \b}{\cal C}[\W t,n-1,\b_{\W t},n|2,n-1,\b_2,n]\right\}\W
A_{n-1}(2,n-1,\b_2,n).\Label{Need-to-proof-2}\eea

For the first term in \eqref{Need-to-proof-2}, if we have the following identity
\bea \boxed{\sum_{\sigma_{\W t},\b_{\W t}\in S_{n-4}}
{\cal D}[t,\sigma_t,n-1,n|\W t,\sigma_{\W t},n-1,n] {\cal
S}[\sigma_{\W t}|\b_{\W t}]_{p_{n-1}}{\cal C}[\W t,n-1,\b_{\W t},n|t,n-1,\b_t,n]= {\cal
S}[\sigma_t|\b_t]_{p_{n-1}}},~~\Label{KLT-equiv}\eea
the first term can be simplified as
\bea & & \sum_{\sigma_2,\b_2\in S_{n-4}}
{\W\la_1^{\dot \a}\W\la_1^{\dot \b}  \Spaa{n|n-1}\over \Spaa{n-1|1}
\Spaa{n|1}}A_{n-1}(2,\sigma_2,n-1,n) \sum_{\W t=2}^{n-2}{\cal
S}[\sigma_2|\b_2]_{p_{n-1}} \left\{J_{\W t,\dot \a \dot \b}\W
A_{n-1}(2,n-1,\b_2,n)\right\}\nn
& = & \sum_{\sigma_2,\b_2\in S_{n-4}}
{\W\la_1^{\dot \a}\W\la_1^{\dot \b}  \Spaa{n|n-1}\over \Spaa{n-1|1}
\Spaa{n|1}}A_{n-1}(2,\sigma_2,n-1,n) {\cal
S}[\sigma_2|\b_2]_{p_{n-1}}\left\{ \left[\sum_{\W t=2}^{n-2}J_{\W t,\dot \a \dot \b}\right]\W
A_{n-1}(2,n-1,\b_2,n)\right\}\nn
& = & 0,\eea
where we have used angular momentum conservation
\bea & & \W\la_1^{\dot \a}\W\la_1^{\dot \b}\left\{ \sum_{\W t=2}^{n-2}J_{\W t,\dot \a \dot \b}\right\}\left\{\W
A_{n-1}(2,n-1,\b_2,n)\right\}=   \W\la_1^{\dot \a}\W\la_1^{\dot \b}\left\{ \sum_{\W t=2}^{n}J_{\W t,\dot \a \dot \b}\right\}\left\{\W
A_{n-1}(2,n-1,\b_2,n)\right\}\nn
&= & (-\W\la_1^{\dot \a}\W\la_1^{\dot \b} J_{\W t=1,\dot \a \dot \b})\left\{\W
A_{n-1}(2,n-1,\b_2,n)\right\}=0~.\eea

For the second term in \eqref{Need-to-proof-2}, if we have the following identity
\bea \boxed{0= \sum_{\W t=2}^{n-2}\sum_{\sigma_{\W t},\b_{\W t}\in S_{n-4}}
{\cal D}[t,\sigma_t,n-1,n|\W t,\sigma_{\W t},n-1,n] {\cal
S}[\sigma_{\W t}|\b_{\W t}]_{p_{n-1}}   J_{\W t,\dot \a \dot \b}\left\{{\cal C}[\W t,n-1,\b_{\W t},n|t,n-1,\b_t,n]\right\}},~~~~\Label{non-trivial}\eea
for arbitrary $t\in \{2,3,...,n-2\}$ and related  $\{\sigma_t, \b_t\}$, the contribution vanishes also.

Identities \eref{KLT-equiv} and \eref{non-trivial} are the consistency requirement of the new soft
graviton theorem and the old KLT formula. While the first identity can be understood from the changing of the basis (we will discuss it shortly), the second identity is very nontrivial. Currently, we do not have  an analytic proof for them although in our few examples, we have  checked them explicitly. We believe the knowledge of these two identities will tell us some important aspects of momentum kernel ${\cal S}[\a|\b]$.

Now we present the physical understanding of the first identity  \eref{KLT-equiv}.
Noticing that we have many $(n-3)!$ symmetry KLT forms. They are equivalent to each other, but it is hard to see that from the angle of BCJ relation for color-ordered Yang-Mills theory. In other words, we have
\bea { M}_{n-1} & = &  \sum_{\sigma_t,\b_t\in S_{n-4}}
A_{n-1}(t,\sigma_t,n-1,n) {\cal
S}[\sigma_t|\b_t]_{p_{n-1}} \W
A_{n-1}(t,n-1,\b_t,n)\nn
& = &  \sum_{\sigma_{\W t},\b_{\W t}\in S_{n-4}}
A_{n-1}(\W t,\sigma_{\W t},n-1,n) {\cal
S}[\sigma_{\W t}|\b_{\W t}]_{p_{n-1}} \W
A_{n-1}(\W t,n-1,\b_{\W t},n)\eea
where $\sigma_t,\b_t$ is the set of removing element $t$ from $\{2,3,...,n-2\}$. Plugging the transformation of basis  \eqref{minimal-basis} back, we have
\bea & & \sum_{\sigma_{\W t},\b_{\W t}\in S_{n-4}}
A_{n-1}(\W t,\sigma_{\W t},n-1,n) {\cal
S}[\sigma_{\W t}|\b_{\W t}]_{p_{n-1}} \W
A_{n-1}(\W t,n-1,\b_{\W t},n)\nn
& = & \sum_{\sigma_{\W t},\b_{\W t}\in S_{n-4}}
\left\{\sum_{\sigma_{ t}\in S_{n-4}} A_{n-1}(t,\sigma_t,n-1,n)
{\cal D}[t,\sigma_t,n-1,n|\W t,\sigma_{\W t},n-1,n]\right\} {\cal
S}[\sigma_{\W t}|\b_{\W t}]_{p_{n-1}}\nn  & & \left\{ \sum_{\b_{ t}\in S_{n-4}} {\cal C}[\W t,n-1,\b_{\W t},n|t,n-1,\b_t,n]
\W
A_{n-1}(t,n-1,\b_t,n)\right\}\nn
& = &\sum_{\sigma_{ t}\in S_{n-4}} A_{n-1}(t,\sigma_t,n-1,n)\left\{\sum_{\sigma_{\W t},\b_{\W t}\in S_{n-4}}
{\cal D}[t,\sigma_t,n-1,n|\W t,\sigma_{\W t},n-1,n] {\cal
S}[\sigma_{\W t}|\b_{\W t}]_{p_{n-1}}{\cal C}[\W t,n-1,\b_{\W t},n|t,n-1,\b_t,n]\right\}\nn
& & \W
A_{n-1}(t,n-1,\b_t,n)\eea
Because  the independence of the BCJ basis, we should obtain the identity \eqref{KLT-equiv}.

\subsection{The sub-sub-leading part from KLT relation}

Now we consider the sub-sub-leading order. From the KLT formula, we have
\bea & & ( \eps^{-2} A_{L,0}+\eps^{-1} A_{L,1}+\eps^{0} A_{L,2}+..) (\eps {\cal S}_0+\eps^2 {\cal S}_1+\eps^3 {\cal S}_2+...)( \eps^{-2} A_{R,0}+\eps^{-1} A_{R,1}+\eps^{0} A_{R,2}+..)\nn
& = & \eps^{-3}  A_{L,0} {\cal S}_0 A_{R,0} +\eps^{-2} (A_{L,1} {\cal S}_0 A_{R,0}+ A_{L,0} {\cal S}_1 A_{R,0}+A_{L,0} {\cal S}_0 A_{R,1})\nn
& + & \eps^{-1} (A_{L,2} {\cal S}_0 A_{R,0}+ A_{L,0} {\cal S}_2 A_{R,0}+A_{L,0} {\cal S}_0 A_{R,2}
+A_{L,1} {\cal S}_1 A_{R,0}+ A_{L,0} {\cal S}_1 A_{R,1}+A_{L,1} {\cal S}_0 A_{R,1})+...\eea
Thus we see that to use this formula to study the sub-sub-leading singularity, we need to get the information of $\eps^{0} A_{L,2}$, which does not have the universal structure and has not been fully discussed.

\section{Examples  }

Having the general frame in previous section, we will present a few examples to demonstrate our ideas. In
this section, we will give examples of $n=5,6$ while the more complicated example of $n=7$ will be given in the Appendix.

\subsection{The case  $n=5$}
Following our  convention, in the stripped amplitude,
$\tilde{\lambda}_{4}$ and $\tilde{\lambda}_{5}$  should be replaced by
\bea
\tilde{\lambda}_{4}=-\sum_{k=2,3}\frac{\left\langle 5|k\right\rangle }{\left\langle 5|4\right\rangle }\tilde{\lambda}_{k}-\epsilon\frac{\left\langle 5|1\right\rangle }{\left\langle 5|4\right\rangle }\tilde{\lambda}_{1},\,\tilde{\lambda}_{5}=-\sum_{k=2,3}\frac{\left\langle 4|k\right\rangle }{\left\langle 4|5\right\rangle }\tilde{\lambda}_{k}-\epsilon\frac{\left\langle 4|1\right\rangle }{\left\langle 4|5\right\rangle }\tilde{\lambda}_{1}.
\eea
In particular that $\frac{d}{d\tilde{\lambda}_{4}^{\dot{\beta}}}\tilde{A}=\frac{d}{d\tilde{\lambda}_{5}^{\dot{\beta}}}\tilde{A}=0$,
and therefore $J_{4\dot{\alpha}\dot{\beta}}\tilde{A}=J_{5\dot{\alpha}\dot{\beta}}\tilde{A}=0$.
At $5$-points it is relatively straightforward to write down all
of the terms in $\Delta$ as
\bea & & \Delta_{n=5}  = {\W\la_1^{\dot \a}\W\la_1^{\dot \b}  \Spaa{5|4}\over \Spaa{4|1}
\Spaa{5|1}} \left\{ A_4(2,3,4,5) {\cal S}[3|3]_{p_4} \left(J_{2,\dot \a \dot \b}\W
A_4(2,4,3,5) \right)  +A_4(3,2,4,5) {\cal S}[2|2]_{p_4} \left(J_{3,\dot \a \dot \b}\W
A_4(3,4,2,5) \right)\right\}\nn
& = &{\W\la_1^{\dot \a}\W\la_1^{\dot \b}  \Spaa{5|4}\over \Spaa{4|1}
\Spaa{5|1}} \left\{ A_4(2,3,4,5) s_{3 4} \left(J_{2,\dot \a \dot \b}\W
A_4(2,4,3,5) \right)  +A_4(3,2,4,5) s_{2 4} \left(J_{3,\dot \a \dot \b}\W
A_4(3,4,2,5) \right)\right\} \Label{delta-5}\eea
Now we do the changing of basis, i.e., using the BCJ relation to write
\bea A_4(3,2,4,5) & = & { s_{34}\over s_{24} } A_4(2,3,4, 5)\nn
\tilde{A}_{4}(3,4,2,5) & = & (-)^{4}\tilde{A}_{4}(5,2,4,3)=\tilde{A}_{4}(2,4,3,5)\eea
Plugging them back we get
\bea \Delta_{n=5}={\W\la_1^{\dot \a}\W\la_1^{\dot \b}  \Spaa{5|4}\over \Spaa{4|1}
\Spaa{5|1}} A_4(3,2,4,5) s_{2\WH 4} \left\{\left(J_{3,\dot \a \dot \b}+J_{2,\dot \a \dot \b}\right)\W
A_4(3,4,2,5) \right\}=0 \eea
by  angular momentum conservation $\sum_{i=2}^5 J_i \W A=0$
(where $J_4 \W A=J_5 \W A=0$ has been used). For this case, two identities \eref{KLT-equiv} and \eref{non-trivial} are trivial to check.

\subsection{The case $n=6$}

 For $n=6$ the difference term $\Delta_{n=6}$ splits into
three parts: $t=2$, $3$ and $4$,
\bea \Delta_{n=6} & = & \Delta_{n=6}^{t=2}+ \Delta_{n=6}^{t=3}+\Delta_{n=6}^{t=4}
\eea
where we have solved
\bea
\tilde{\lambda}_{5}=-\sum_{k=2,3,4}\frac{\left\langle 6|k\right\rangle }{\left\langle 6|5\right\rangle }\tilde{\lambda}_{k}-\epsilon\frac{\left\langle 6|1\right\rangle }{\left\langle 6|5\right\rangle }\tilde{\lambda}_{1},\hspace{1cm}\tilde{\lambda}_{6}=-\sum_{k=2,3,4}\frac{\left\langle 5|k\right\rangle }{\left\langle 5|6\right\rangle }\tilde{\lambda}_{k}-\epsilon\frac{\left\langle 5|1\right\rangle }{\left\langle 5|6\right\rangle }\tilde{\lambda}_{1}.
 \Label{6pt-mom-conserv-cond}
\eea
For simplicity in the following discussion we suppress a common factor
$(-)^{n+1}\frac{\left\langle n|n-1\right\rangle }{\left\langle n-1|1\right\rangle \left\langle n|1\right\rangle }\tilde{\lambda}_{1}^{\dot{\alpha}}\tilde{\lambda}_{1}^{\dot{\beta}}$
from the difference terms, thus we can write
\bea
%
 \Delta_{n=6}^{t=2} & = & A_5(2,3,4,5,6) {\cal S}[3,4|3,4]_{p_5} ( J_2
\W A_5(2,5,3,4,6))+  A_5(2,3,4,5,6) {\cal S}[3,4|4,3]_{p_5} ( J_2 \W
A_5(2,5,4,3,6)) \nn
& + & A_5(2,4,3,5,6) {\cal S}[4,3|3,4]_{p_5} ( J_2 \W
A_5(2,5,3,4,6))+  A_5(2,4,3,5,6) {\cal S}[4,3|4,3]_{p_5} ( J_2 \W
A_5(2,5,4,3,6))\nn
\Delta_{n=6}^{t=3}& = & A_5(3,2,4,5,6) {\cal S}[2,4|2,4]_{p_5}(J_3 \W A_5(3,5,2,4,6))+
A_5(3,2,4,5,6) {\cal S}[2,4|4,2]_{p_5}(J_3 \W A_5(3,5,4,2,6))\nn
& + & A_5(3,4,2,5,6) {\cal S}[4,2|2,4]_{p_5}(J_3 \W A_5(3,5,2,4,6))+
A_5(3,4,2,5,6) {\cal S}[4,2|4,2]_{p_5}(J_3 \W A_5(3,5,4,2,6))\nn
\Delta_{n=6}^{t=4}& = & A_5(4,2,3,5,6) {\cal S}[2,3|2,3]_{p_5} ( J_4 \W
A_5(4,5,2,3,6))+ A_5(4,2,3,5,6) {\cal S}[2,3|3,2]_{p_5} ( J_4 \W
A_5(4,5,3,2,6)) \nn
& + & A_5(4,3,2,5,6) {\cal S}[3,2|2,3]_{p_5} ( J_4 \W
A_5(4,5,2,3,6))+ A_5(4,3,2,5,6) {\cal S}[3,2|3,2]_{p_5} ( J_4 \W
A_5(4,5,3,2,6)) \nn\eea
Now we translate all amplitudes $A$ into the basis $\{A(6,2,4,3,5), A(6,2,3,4,5)\}$
\bea A_5(6,4,2,3,5) & = & { (s_{43}+s_{45}) A_5(6,2,4,3,5)+s_{45} A_5(6,2,3,4,5) \over s_{46}}\nn
A_5(6,3,2,4,5) & = & { (s_{34}+s_{35}) A_5(6,2,3,4,5)+s_{35} A_5(6,2,4,3,5) \over s_{36}}\nn
A_5(6,4,3,2,5) & = & { -s_{24} s_{35} A_5(6,2,4,3,5)- s_{45}(s_{25}+s_{23}) A_5(6,2,3,4,5) \over s_{46}s_{25}}\nn
 A_5(6,3,4,2,5) & = & { -s_{23} s_{45} A_5(6,2,3,4,5)- s_{35}(s_{25}+s_{24}) A_5(6,2,4,3,5) \over s_{36}s_{25}}\eea
and all amplitudes $\W A_5$ into the basis $\{\W A_5 (2,5,3,4,6) , \W A_5(2,5,4,3,6)\}$
\bea  \W A_5(3,5,2,4,6) & = & {- \W A_5 (2,5,3,4,6) (s_{45}+s_{43})- \W A_5(2,5,4,3,6) s_{45}\over s_{24}} \nn
\W A_5(4,5,2,3,6) & = & {- \W A_5 (2,5,4,3,6) (s_{35}+s_{43})- \W A_5(2,5,3,4,6) s_{35}\over s_{23}} \nn
\W A_5(3,5,4,2,6)) & = & { -(s_{43}+s_{46})\W A_5(2,5,4,3,6) -s_{46} \W A_5 (2,5,3,4,6)\over s_{24}}\nn
\W A_5(4,5,3,2,6)) & = & { -(s_{43}+s_{36})\W A_5(2,5,3,4,6) -s_{36} \W A_5 (2,5,4,3,6)\over s_{23}}\eea
Putting it back with some calculation we have
\bea \Delta_{n=6}^{t=3} & = & A_5(2,3,4,5,6)\left\{ -s_{45} (s_{23}+s_{25}) (J_{3,\dot\a\dot \b} {- \W A_5 (2,5,3,4,6) (s_{45}+s_{43})- \W A_5(2,5,4,3,6) s_{45}\over s_{24}})\right.\nn & & \left. +s_{45} s_{26} (J_{3,\dot\a\dot \b} { -(s_{43}+s_{46})\W A_5(2,5,4,3,6) -s_{46} \W A_5 (2,5,3,4,6)\over s_{24}})\right\} \nn
& +& A_5(2,4,3,5,6)\left\{ -s_{35} s_{24}(J_{3,\dot\a\dot \b}{- \W A_5 (2,5,3,4,6) (s_{45}+s_{43})- \W A_5(2,5,4,3,6) s_{45}\over s_{24}}) \right\}\eea
Further simplification by using ( notice that $J_{3,\dot\a\dot \b} s_{24}=0$)
\bea & &  (s_{23}+s_{25}) (J_{3,\dot\a\dot \b} (s_{45}+s_{43}))- s_{26} (J_{3,\dot\a\dot \b} s_{46})=s_{24} (J_{3,\dot\a\dot \b} s_{46})\nn
& &   (s_{23}+s_{25}) (J_{3,\dot\a\dot \b} s_{45})- s_{26} (J_{3,\dot\a\dot \b} (s_{43}+s_{46}))=- s_{24} (J_{3,\dot\a\dot \b} s_{45}) \eea
leads
\bea & & \Delta_{n=6}^{t=3}  =  A_5(2,3,4,5,6)\left\{ {\cal S}[3,4|3,4]_{p_5} ( J_{3,\dot\a\dot \b}
\W A_5(2,5,3,4,6))+   {\cal S}[3,4|4,3]_{p_5} ( J_{3,\dot\a\dot \b} \W
A_5(2,5,4,3,6))\right. \nn & &\left. +{s_{45}(J_{3,\dot\a\dot \b} s_{46})\W A_5 (2,5,3,4,6)}- {s_{45}\W A_5 (2,5,4,3,6)(J_{3,\dot\a\dot \b} s_{45})}\right\} \nn
& +& A_5(2,4,3,5,6)\left\{ {\cal S}[4,3|3,4]_{p_5} ( J_{3,\dot\a\dot \b} \W
A_5(2,5,3,4,6))+  {\cal S}[4,3|4,3]_{p_5} ( J_{3,\dot\a\dot \b} \W
A_5(2,5,4,3,6))\right. \nn & & \left.- s_{35}   \W A_5 (2,5,3,4,6)(J_{3,\dot\a\dot \b} s_{46})+s_{35} \W A_5(2,5,4,3,6) (J_{3,\dot\a\dot \b} s_{45}) \right\}\eea
Notice that part of them (i.e., the part with $J$ acting only on $\W A$) is exactly the same as  $\Delta_{n=6}^{t=2}$ except the $J_{2,\dot\a \dot \b}$ is replaced by $J_{3,\dot\a\dot \b}$. It is nothing, but the explicit checking the  identity \eref{KLT-equiv} with $t=2, \W t=3$.

Doing  similar calculation we found
\bea  \Delta_{n=6}^{t=4}
& = &   A_5(2,3,4,5,6)\left\{ {\cal S}[3,4|3,4]_{p_5} ( J_{4,\dot \a \dot \b}
\W A_5(2,5,3,4,6))+   {\cal S}[3,4|4,3]_{p_5} ( J_{4,\dot \a \dot \b} \W
A_5(2,5,4,3,6))\right. \nn & &\left. +{s_{45}(J_{4,\dot \a \dot \b} s_{35})\W A_5 (2,5,3,4,6)}+ {s_{45}\W A_5 (2,5,4,3,6)(J_{4,\dot \a \dot \b}( s_{34}+s_{35}))}\right\} \nn
& +& A_5(2,4,3,5,6)\left\{ {\cal S}[4,3|3,4]_{p_5} ( J_{4,\dot \a \dot \b} \W
A_5(2,5,3,4,6))+  {\cal S}[4,3|4,3]_{p_5} ( J_{4,\dot \a \dot \b} \W
A_5(2,5,4,3,6))\right. \nn & & \left.- s_{35}   \W A_5 (2,5,3,4,6)(J_{4,\dot \a \dot \b} s_{35})+s_{35} \W A_5(2,5,4,3,6) (J_{4,\dot \a \dot \b} s_{36}) \right\}\eea
where again the identity \eref{KLT-equiv} with $t=2, \W t=4$ has been checked. Thus when we sum up three terms $\Delta_{n=6}^{t=2}, \Delta_{n=6}^{t=3},\Delta_{n=6}^{t=4}$, the part with $J$ acting directly on $\W A$ vanishes by angular momentum conservation and we are left with
\bea R & =& A_5(2,3,4,5,6)\left\{ +{s_{45}(J_{3,\dot \a \dot \b} s_{46})\W A_5 (2,5,3,4,6)}- {s_{45}\W A_5 (2,5,4,3,6)(J_{3,\dot \a \dot \b} s_{45})}\right. \nn & &\left. +{s_{45}(J_{4,\dot \a \dot \b} s_{35})\W A_5 (2,5,3,4,6)}- {s_{45}\W A_5 (2,5,4,3,6)(J_{4,\dot \a \dot \b} s_{36})}\right\} \nn
& +& A_5(2,4,3,5,6)\left\{ - s_{35}   \W A_5 (2,5,3,4,6)(J_{3,\dot \a \dot \b} s_{46})+s_{35} \W A_5(2,5,4,3,6) (J_{3,\dot \a \dot \b} s_{45})\right. \nn & & \left.- s_{35}   \W A_5 (2,5,3,4,6)(J_{4,\dot \a \dot \b} s_{35})+s_{35} \W A_5(2,5,4,3,6) (J_{4,\dot \a \dot \b} s_{36}) \right\}\eea
where $J$ acts only on $s_{ij}$. Using
\bean J_{3,\dot \a \dot \b} s_{i5}=-\Spaa{5|i} {\Spaa{6|3}\over \Spaa{6|5}}\W \la_{3,\dot \a} \W\la_{i, \dot \b},
~~~~J_{3,\dot \a \dot \b} s_{i6}=-\Spaa{6|i} {\Spaa{5|3}\over \Spaa{5|6}}\W \la_{3,\dot \a} \W\la_{i, \dot \b},~~~  J_{3,\dot \a \dot \b} s_{i3}=
+\Spaa{3|i}\W \la_{3,\dot \a} \W\la_{i, \dot \b}   ,~~~~~~i=2,4\nn
J_{4,\dot \a \dot \b} s_{i5}=-\Spaa{5|i} {\Spaa{6|4}\over \Spaa{6|5}}\W \la_{4,\dot \a} \W\la_{i, \dot \b},
~~~~J_{4,\dot \a \dot \b} s_{i6}=-\Spaa{6|i} {\Spaa{5|4}\over \Spaa{5|6}}\W \la_{4,\dot \a} \W\la_{i, \dot \b},~~~  J_{4,\dot \a \dot \b} s_{i4}=
+\Spaa{4|i}\W \la_{4,\dot \a} \W\la_{i, \dot \b}   ,~~~~~~i=2,3\eean
we see immediately that $R=0$. In other words, we have explicitly checked the second identity \eref{non-trivial} for the special case.

\section{Conclusion}
In this paper, we studied the new soft graviton theorem from the angle of  KLT relation. We have demonstrated that how the new soft gluon theorem are glued together by KLT formula to produce the corresponding soft
theorem for gravity. In the process, two important identities \eref{KLT-equiv} and \eref{non-trivial}
has been observed.

There are a lot of open questions deserve to be investigated. First, the two identities need an analytic proof. Secondly, the  sub-sub-leading soft factor in KLT relation should be understood. Although at this order, contributions from non-universal soft part of color ordered Yang-Mills amplitudes appear, we guess that their effects will be canceled out by nice property of momentum kernel ${\cal S}$. It will be fascinating to see how it happens. Thirdly, in this paper, we have focused on the 4D, it will be interesting to discuss it in general dimension since KLT formula holds in general dimension. Finally, there are also other general formulas  for gravity amplitudes (such as these given in \cite{Cachazo:2012da, Cachazo:2012kg, Cachazo:2013zc} ) and it will be nice to see how the new soft graviton theorem makes its appearance.

\subsection*{Acknowledgements}
Y. J. Du would like to acknowledge the EU programme Erasmus Mundus Action 2, Project 9  and
the International Postdoctoral Exchange Fellowship Program of China for supporting his postdoctoral research in Lund University (with Fudan University as the home university).
Y. J. Du's research is supported in parts by the NSF of China Grant No.11105118, China Postdoctoral Science Foundation No.2013M530175 and the Fundamental Research Funds for the Central Universities of Fudan University No.20520133169.
C. F. is grateful for Y-T Huang for the helpful discussions.
The work of C. F. was supported from National Science Council, 50 billions project of
Ministry of Education and National Center for Theoretical Science,
Taiwan, Republic of China as well as the support from S.T. Yau
center of National Chiao Tung University. 
Y.W. would like acknowledge the support from NSF grant PHY-1316617.
B.F is supported, in part,
by fund from Qiu-Shi and Chinese NSF funding under contract
No.11031005, No.11135006, No. 11125523.

\appendix

\section{Example with $n=7$}
In this appendix we verify that identities (\ref{KLT-equiv}) and (\ref{non-trivial})
are holding at $n=7$.
Our strategy used in previous examples applies to $7$-points, although
the complexity involved increases considerably.
As in the previous examples, we choose to work in a convenient minimum
basis $\tilde{A}(2,6,\beta_{2},7),A(2,\sigma_{2},6,7) $ (Here we use $A$ instead of $A_6$ for short), i.e., do the following transformation with $t=3,4,5$:
\begin{eqnarray}
\tilde{A}(t,6,\beta_{t},7) & = & \sum_{\beta_{t}\in S_{3}}\mathcal{C}\left[t,6,\beta_{t},7|2,6,\beta_{2},7\right]\tilde{A}(2,6,\beta_{2},7),~~~\Label{n=7-A1}\\
A(t,\sigma_{t},6,7) & = & \sum_{\sigma_{t}\in S_{3}}
A(2,\sigma_{2},6,7)
\mathcal{D}\left[2,\sigma_{2},6,7 | t,\sigma,6,7 \right].\nonumber
\end{eqnarray}
Our task then amounts to showing that, 
for both identities,
terms associated with each independent
product of basis amplitudes $A\tilde{A}$ match accordingly
for both sides of the equations (\ref{KLT-equiv}) and (\ref{non-trivial}). In the
discussion below we focus on terms containing $\tilde{A}(2,6,3,4,5,7)$,
namely when $\beta_{2}=\{3,4,5\}$. The rest of the coefficients follow
similar argument up to permutations of $\{3,4,5\}$. In principle
it is straightforward to work out all translation coefficients $\mathcal{C}$,
$\mathcal{D}$ and check if the identities are holding.
However we can perform the calculation in a slightly more organized
manner. In particular note that common factors are quite often shared between
different translation coefficients.

For the purpose of demonstration  
let us consider translating a specific
amplitude $\tilde{A}(3,6,2,4,5,7)$ into minimum basis. This can be
done by first expressing the amplitude in the $\tilde{A}(2,\dots,7)$
Kleiss-Klein (KK) basis that fixes legs $2$ and $7$ at both ends,
and then subsequently translating to the $\tilde{A}(2,6,\dots,7)$
minimum basis of interest where legs  $6$ and $2$  are adjacent:
\begin{eqnarray}
\tilde{A}(3,6,2,4,5,7) & = & \tilde{A}(2,4,5,6,3,7)+\tilde{A}(2,4,6,5,3,7)+\tilde{A}(2,4,6,3,5,7)+\tilde{A}(2,6,4,5,3,7)\\
 &  & +\tilde{A}(2,6,4,3,5,7)+\tilde{A}(2,6,3,4,5,7)\nonumber \\
 & = & \left(1-\,\frac{(s_{42}+s_{46}+s_{43})}{s_{42}}+E\left[45,3|345\right]\right)\tilde{A}(2,6,3,4,5,7)\nonumber \\
 &  & +\dots\left(\text{terms not contributing to }\tilde{A}(2,6,3,4,5,7)\right),\nonumber
\end{eqnarray}
where in the third line we used BCJ relation to remove the ill-favored
leg $4$ between $2$ and $6$ in the next to adjacent amplitude $\tilde{A}(2,4,6,5,3,7)$,
and we introduced the shorthand notation $E\left[45,3|345\right]$
to denote the next-to-next-to adjacent expansion coefficient,
\begin{equation}
\tilde{A}(2,\{4,5\},6,\{3\},7)=\sum_{\sigma}E\left[45,3|\sigma\right]\tilde{A}(2,6,\sigma,7).
\end{equation}
The coefficient $E\left[45,3|345\right]$ can be determined from simultaneous
equations consisting of BCJ relations, yielding
\begin{eqnarray}
E\left[45,3|345\right] & = & \left.\frac{(-1)}{s_{42}s_{52}-(s_{42}+s_{45})(s_{52}+s_{54})}\right[-(s_{42}+s_{45}+s_{46}+s_{43})(s_{52}+s_{56}+s_{53}+s_{54})\\
 &  & \left.+\frac{(s_{42}+s_{45})(s_{52}+s_{54}+s_{56}+s_{53})(s_{42}+s_{46}+s_{43})}{s_{42}}\right].\nonumber
\end{eqnarray}
All translation coefficients can be determined via similar procedures.
Explicitly we have, for the $t=3$ sector,
\begin{eqnarray}
\mathcal{C}\left[3,6,2,4,5,7|2,6,3,4,5,7\right] & = & 1-\,\frac{(s_{42}+s_{46}+s_{43})}{s_{42}}+E\left[45,3\right]\label{eq:transltn-c1}\\
\mathcal{C}\left[3,6,2,5,4,7|2,6,3,4,5,7\right] & = & -\,\frac{(s_{52}+s_{56}+s_{53}+s_{54})}{s_{52}}+E\left[54,3\right]\nonumber \\
\mathcal{C}\left[3,6,4,2,5,7|2,6,3,4,5,7\right] & = & \frac{(s_{42}+s_{46}+s_{43})}{s_{42}}-E\left[45,3\right]-E\left[54,3\right]\nonumber\\
\mathcal{C}\left[3,6,4,5,2,7|2,6,3,4,5,7\right] & = & E\left[54,3\right]\nonumber\\
\mathcal{C}\left[3,6,5,2,4,7|2,6,3,4,5,7\right] & = & \frac{(s_{52}+s_{56}+s_{53}+s_{54})}{s_{52}}-E\left[45,3\right]-E\left[54,3\right]\nonumber\\
\mathcal{C}\left[3,6,5,4,2,7|2,6,3,4,5,7\right] & = & E\left[45,3\right].\label{eq:transltn-c6}
\end{eqnarray}
For $t=4$ we have
\begin{eqnarray}
\mathcal{C}\left[4,6,2,3,5,7|2,6,3,4,5,7\right] & = & 1-\,\frac{(s_{32}+s_{36})}{s_{32}}+E\left[35,4\right]\\
\mathcal{C}\left[4,6,2,5,3,7|2,6,3,4,5,7\right] & = & -\,\frac{(s_{52}+s_{56}+s_{53}+s_{54})}{s_{52}}+E\left[53,4\right]\nonumber \\
\mathcal{C}\left[4,6,3,2,5,7|2,6,3,4,5,7\right] & = & \frac{(s_{32}+s_{36})}{s_{32}}-E\left[35,4\right]-E\left[53,4\right]\nonumber\\
\mathcal{C}\left[4,6,3,5,2,7|2,6,3,4,5,7\right] & = & E\left[53,4\right]\nonumber\\
\mathcal{C}\left[4,6,5,2,3,7|2,6,3,4,5,7\right] & = & \frac{(s_{52}+s_{56}+s_{53}+s_{54})}{s_{52}}-E\left[35,4\right]-E\left[53,4\right]\nonumber\\
\mathcal{C}\left[4,6,5,3,2,7|2,6,3,4,5,7\right] & = & E\left[35,4\right],\nonumber
\end{eqnarray}
and similarly for $t=5$,
\begin{eqnarray}
\mathcal{C}\left[5,6,2,3,4,7|2,6,3,4,5,7\right] & = & 1-\,\frac{(s_{32}+s_{36})}{s_{32}}+E\left[34,5\right]\\
\mathcal{C}\left[5,6,2,4,3,7|2,6,3,4,5,7\right] & = & -\,\frac{(s_{42}+s_{46}+s_{43})}{s_{42}}+E\left[43,5\right]\nonumber \\
\mathcal{C}\left[5,6,3,2,4,7|2,6,3,4,5,7\right] & = & \frac{(s_{32}+s_{36})}{s_{32}}-E\left[34,5\right]-E\left[43,5\right]\\
\mathcal{C}\left[5,6,3,4,2,7|2,6,3,4,5,7\right] & = & E\left[43,5\right]\\
\mathcal{C}\left[5,6,4,2,3,7|2,6,3,4,5,7\right] & = & \frac{(s_{42}+s_{46}+s_{43})}{s_{42}}-E\left[34,5\right]-E\left[43,5\right]\\
\mathcal{C}\left[5,6,4,3,2,7|2,6,3,4,5,7\right] & = & E\left[34,5\right],
\end{eqnarray}
whereas the next-to-next-to adjacent expansion coefficients are given
by
\begin{eqnarray}
E\left[54,3\right] & = & \left.\frac{(-1)}{s_{52}s_{42}-(s_{52}+s_{54})(s_{42}+s_{45})}\right[-(s_{52}+s_{54}+s_{56}+s_{53})(s_{42}+s_{46}+s_{43})\\
 &  & \left.+\frac{(s_{52}+s_{54})(s_{42}+s_{45}+s_{46}+s_{43})(s_{52}+s_{56}+s_{53}+s_{54})}{s_{52}}\right]\nonumber\\
E\left[35,4\right] & = & \left.\frac{(-1)}{s_{32}s_{52}-(s_{32}+s_{35})(s_{52}+s_{53})}\right[-(s_{32}+s_{35}+s_{36})(s_{52}+s_{56}+s_{53}+s_{54})\\
 &  & \left.+\frac{(s_{32}+s_{35})(s_{52}+s_{53}+s_{56}+s_{54})(s_{32}+s_{36})}{s_{32}}\right]\nonumber\\
E\left[53,4\right] & = & \left.\frac{(-1)}{s_{52}s_{32}-(s_{52}+s_{53})(s_{32}+s_{35})}\right[-(s_{52}+s_{53}+s_{56}+s_{54})(s_{32}+s_{36})\\
 &  & \left.+\frac{(s_{52}+s_{53})(s_{32}+s_{35}+s_{36})(s_{52}+s_{56}+s_{53}+s_{54})}{s_{52}}\right]\nonumber\\
E\left[34,5\right] & = &
\left.\frac{(-1)}{s_{32}s_{42}-(s_{32}+s_{34})(s_{42}+s_{43})}\right[-(s_{32}+s_{34}+s_{36})(s_{42}+s_{46}+s_{43})\\
 &  & \left.+\frac{(s_{32}+s_{34})(s_{42}+s_{43}+s_{46})(s_{32}+s_{36})}{s_{32}}\right]\nonumber
  \\
E\left[43,5\right] & = &
 \left.\frac{(-1)}{s_{42}s_{32}-(s_{42}+s_{43})(s_{32}+s_{34})}\right[-(s_{42}+s_{43}+s_{46})(s_{32}+s_{36})\\
 &  & \left.+\frac{(s_{42}+s_{43})(s_{32}+s_{34}+s_{36})(s_{42}+s_{46}+s_{43})}{s_{42}}\right]\nonumber
\end{eqnarray}

\begin{flushleft}
\textbf{Identity (\ref{KLT-equiv}) }
 \end{flushleft}
Let us first verify identity (\ref{KLT-equiv}) for the case when $\beta_{2}=\{345\}$, or
equivalently that
\begin{equation}
\sum_{\sigma_{\tilde{t}},\beta_{\tilde{t}}}\, A(\tilde{t},\sigma_{\tilde{t}},67)\mathcal{S}\left[\sigma_{\tilde{t}}|\beta_{\tilde{t}}\right]\mathcal{C}\left[\tilde{t},6,\beta_{\tilde{t}},7\right]=\sum_{\sigma_{2}\in perm\{345\}}A(2,\sigma_{2},67)\mathcal{S}\left[\sigma_{2}|345\right]_{6}\label{eq:identity-415}
\end{equation}
for each $\tilde{t}$. To keep the derivation simple we introduce
the following shorthand notation for repeatedly occurring factors
\begin{equation}
T_{\tilde{t}}(\beta_{\tilde{t}})=\sum_{\sigma_{\tilde{t}}}\, A(\tilde{t},\sigma_{\tilde{t}},67)\mathcal{S}\left[\sigma_{\tilde{t}}|\beta_{\tilde{t}}\right].
\end{equation}
so that for example when $\tilde{t}=5$, equation (\ref{eq:identity-415})
reads
\begin{equation}
\sum_{\beta_{5}\in perm\{234\}}\, T_{5}(\beta_{5})\mathcal{C}\left[5,6,\beta_{5},7\right]=T_{2}(345).\label{eq:t-tilde-5}
\end{equation}
Substituting the explicit expressions for translation coefficients
$\mathcal{C}$s, the left hand side of the above equation becomes
\begin{eqnarray}
 &  & T_{5}(234)+\left(-T_{5}(234)+T_{5}(324)\right)\frac{s_{32}+s_{36}}{s_{32}}\\
 &  & +\left(-T_{5}(243)+T_{5}(423)\right)\frac{s_{42}+S_{46}+s_{43}}{s_{42}}\nonumber\\
 &  & +\left(T_{5}(234)-T_{5}(324)-T_{5}(423)+T_{5}(432)\right)E[34,5]\nonumber\\
 &  & +\left(T_{5}(243)-T_{5}(324)+T_{5}(342)-T_{5}(423)\right)E[43,5].\nonumber
\end{eqnarray}
With a little bit more effort, we find that the left hand side of
(\ref{eq:t-tilde-5}) boils down to the following linear combination
of amplitudes.
\begin{eqnarray}
 &  & -s_{36}s_{46}(s_{56}+s_{54}+s_{53}+s_{52})\, A(523467)\label{eq:two-amplitude-sum}\\
 &  & -s_{36}(s_{34}+s_{46})(s_{56}+s_{54}+s_{53}+s_{52})\, A(524367).\nonumber
\end{eqnarray}
On the other hand the right hand side of (\ref{eq:t-tilde-5}) reads
\begin{eqnarray}
T_{2}(345) & = & s_{36}s_{46}s_{56}A(234567)+s_{36}s_{46}(s_{46}+s_{56})A(235467)+s_{36}s_{46}(s_{53}+s_{54}+s_{56})
A(253467)\nn
 &  & +s_{36}(s_{43}+s_{46})s_{56}A(243567)+s_{36}(s_{43}+s_{46})(s_{56}+s_{53})A(245367)\nonumber\\
 &  & +s_{36}(s_{43}+s_{46})(s_{53}+s_{54}+s_{56})A(254367),~~~\label{eq:m2-345}
\end{eqnarray}
We see that the first   line of (\ref{eq:two-amplitude-sum})
matches the sum of the first three terms of equation (\ref{eq:m2-345}),
and similarly the second line of (\ref{eq:two-amplitude-sum})
matches the sum of the last three terms of  (\ref{eq:m2-345})
 because
of BCJ relation, thereby proving the identity (\ref{eq:t-tilde-5}).
The situations when $\tilde{t}=3$ and $4$ can be proved in a likewise
manner.

\begin{flushleft}
\textbf{Identity (\ref{non-trivial}) }
\end{flushleft}

At $7$-points the
difference term $\Delta_{n=7}$ splits into four parts, $\Delta_{n=7}=\sum_{t=2,3,4,5}\,\Delta_{n=7}^{t}$,
where
\begin{equation}
\Delta_{n=7}^{t}=\sum_{\sigma,\beta\in S_{3}}\, A(t,\sigma,6,7)\mathcal{S}[\sigma|\beta]_{6}J_{t}\tilde{A}(t,6,\beta,7)\label{eq:identity}
\end{equation}
Substituting the above expressions into equation (\ref{eq:identity})
and collecting terms, we find as in the previous examples that terms
where angular momentum operate on basis amplitudes $\tilde{A}$ add
up to zero because of angular momentum conservation $\sum_{t}\, J_{t}\tilde{A}(2,6,3,4,5,7)=0$,
leaving us with the collection of terms that $J_{t}$ operate on expansion
coefficients $\mathcal{C}$, which are functions of kinematic variables.
Contributions from the three respective sectors are given by
\begin{eqnarray}
\Delta_{t=3} & = & \left(-T_{3}(245)+T_{3}(425)\right)J_{3}\left(\frac{s_{42}+s_{46}+s_{43}}{s_{42}}\right)+
\left(-T_{3}(254)+T_{3}(524)\right)J_{3}\left(\frac{s_{52}+s_{56}+s_{53}+s_{54}}{s_{52}}\right)
\nn
 &  & +\left(T_{3}(245)-T_{3}(425)-T_{3}(524)+T_{3}(542)\right)J_{3}\left(E\left[45,3\right]\right)\nonumber \\
 &  & +\left(T_{3}(254)-T_{3}(425)+T_{3}(452)-T_{3}(524)\right)J_{3}\left(E\left[54,3\right]\right),\label{eq:t3}
\end{eqnarray}

\begin{eqnarray}
\Delta_{t=4} & = & \left(-T_{3}(235)+T_{3}(325)\right)J_{3}\left(\frac{s_{32}+s_{36}}{s_{32}}\right)+\left(-T_{3}(253)
+T_{3}(523)\right)J_{3}\left(\frac{s_{52}+s_{56}+s_{53}+s_{54}}{s_{52}}\right)\nn
 &  & +\left(T_{3}(235)-T_{3}(325)-T_{3}(523)+T_{3}(532)\right)J_{3}\left(E\left[35,4\right]\right)\nonumber \\
 &  & +\left(T_{3}(253)-T_{3}(325)+T_{3}(4352)-T_{3}(523)\right)J_{3}\left(E\left[53,4\right]\right),\label{eq:t4}
\end{eqnarray}

\begin{eqnarray}
\Delta_{t=5} & = & \left(-T_{3}(234)+T_{3}(324)\right)J_{3}\left(\frac{s_{32}+s_{36}}{s_{32}}\right)+\left(-T_{3}(243)
+T_{3}(423)\right)J_{3}\left(\frac{s_{42}+s_{46}+s_{43}}{s_{42}}\right)\nn
 &  & +\left(T_{3}(234)-T_{3}(324)-T_{3}(423)+T_{3}(432)\right)J_{3}\left(E\left[34,5\right]\right)\nonumber \\
 &  & +\left(T_{3}(243)-T_{3}(324)+T_{3}(342)-T_{3}(423)\right)J_{3}\left(E\left[43,5\right]\right),\label{eq:t5}
\end{eqnarray}

 Generically the operation of $J_{t}$ on kinematic variables must
fall into one of the following categories:
\begin{itemize}
\item $t=3$,
\begin{eqnarray}
J_{3\,\dot{\alpha}\dot{\beta}}s_{i6} & = & \tilde{\lambda}_{3\,(\dot{\alpha}}\tilde{\lambda}_{i\,\dot{\beta})}(-)\frac{\left\langle i6\right\rangle \left\langle 73\right\rangle }{\left\langle 76\right\rangle },\, ~~~~i=2,4,5\label{eq:36}\\
J_{3\,\dot{\alpha}\dot{\beta}}s_{i7} & = & \tilde{\lambda}_{3\,(\dot{\alpha}}\tilde{\lambda}_{i\,\dot{\beta})}(-)\frac{\left\langle i7\right\rangle \left\langle 63\right\rangle }{\left\langle 67\right\rangle }\nonumber \\
J_{3\,\dot{\alpha}\dot{\beta}}s_{i3} & = & \tilde{\lambda}_{3\,(\dot{\alpha}}\tilde{\lambda}_{i\,\dot{\beta})}\left\langle i3\right\rangle \label{eq:34}\\
J_{3\,\dot{\alpha}\dot{\beta}}s_{i\, i^{'}} & = & 0,\,~~~~ i,i^{'}=2,4,5\nonumber
\end{eqnarray}

\item $t=4$,
\begin{eqnarray}
J_{4\,\dot{\alpha}\dot{\beta}}s_{i6} & = & \tilde{\lambda}_{4\,(\dot{\alpha}}\tilde{\lambda}_{i\,\dot{\beta})}(-)\frac{\left\langle i6\right\rangle \left\langle 74\right\rangle }{\left\langle 76\right\rangle },\, ~~~~~i=3,4,5\label{eq:46}\\
J_{4\,\dot{\alpha}\dot{\beta}}s_{i7} & = & \tilde{\lambda}_{4\,(\dot{\alpha}}\tilde{\lambda}_{i\,\dot{\beta})}(-)\frac{\left\langle i7\right\rangle \left\langle 64\right\rangle }{\left\langle 67\right\rangle }\nonumber \\
J_{4\,\dot{\alpha}\dot{\beta}}s_{i4} & = & \tilde{\lambda}_{4\,(\dot{\alpha}}\tilde{\lambda}_{i\,\dot{\beta})}\left\langle i4\right\rangle \nonumber \\
J_{4\,\dot{\alpha}\dot{\beta}}s_{i\, i^{'}} & = & 0,\,~~~~~~ i,i^{'}=3,4,5\nonumber
\end{eqnarray}

\item $t=5$,
\begin{eqnarray}
J_{5\,\dot{\alpha}\dot{\beta}}s_{i6} & = & \tilde{\lambda}_{5\,(\dot{\alpha}}\tilde{\lambda}_{i\,\dot{\beta})}(-)\frac{\left\langle i6\right\rangle \left\langle 75\right\rangle }{\left\langle 76\right\rangle },\,~~~~ i=2,4,5\\
J_{5\,\dot{\alpha}\dot{\beta}}s_{i7} & = & \tilde{\lambda}_{5\,(\dot{\alpha}}\tilde{\lambda}_{i\,\dot{\beta})}(-)\frac{\left\langle i7\right\rangle \left\langle 65\right\rangle }{\left\langle 67\right\rangle }\nonumber \\
J_{5\,\dot{\alpha}\dot{\beta}}s_{i3} & = & \tilde{\lambda}_{5\,(\dot{\alpha}}\tilde{\lambda}_{i\,\dot{\beta})}\left\langle i5\right\rangle \nonumber \\
J_{5\,\dot{\alpha}\dot{\beta}}s_{i\, i^{'}} & = & 0,\,~~~ i,i^{'}=2,4,5\nonumber
\end{eqnarray}

\end{itemize}
Suppose if we are interested in checking terms carrying $\tilde{\lambda}_{3\,(\dot{\alpha}}\tilde{\lambda}_{4\,\dot{\beta})}$.
Before we commence an explicit calculation, note that because all
of the $\mathcal{C}$s do not depend explicitly on leg $7$, from
the list above such a term can only be produced through $J_{3}(s_{46})$,
$J_{3}(s_{43})$, $J_{4}(s_{36})$, $J_{4}(s_{34})$, which allows
us to ignore the $t=5$ sector entirely. Additionally since $s_{34}$
happen to be absent from the $t=4$ translation coefficient $\mathcal{C}$s,
this leaves only $J_{3}(s_{46})$, $J_{3}(s_{43})$, $J_{4}(s_{36})$.
Considering the explicit forms given by equations (\ref{eq:36}),
(\ref{eq:34}) and (\ref{eq:46}) we further note that (again) because
of the absence of the leg $7$ dependence in all $\mathcal{C}$s,
the contributions from $J_{3}(s_{46})$, $J_{3}(s_{43})$, $J_{4}(s_{36})$
together can only cancel through Jacobi identity $\left\langle 43\right\rangle +\frac{\left\langle 73\right\rangle \left\langle 64\right\rangle }{\left\langle 76\right\rangle }+\frac{\left\langle 74\right\rangle \left\langle 36\right\rangle }{\left\langle 76\right\rangle }=0$.
For that to happen, the contribution associated with $J_{3}(s_{46})$,
$J_{3}(s_{43})$, $J_{4}(s_{36})$ must be exactly in the ratio $1:1:-1$,
in other words they must add up to

\begin{equation}
J_{3}(s_{46})X+J_{3}(s_{43})X+J_{4}(s_{36})(-X)=0
\end{equation}
for some factor $X$. In the following discussion we shall see that
indeed this is the case.

First we note that it is relatively easy to confirm that the ratio
between the contributions from $J_{3}(s_{46})$ and $J_{3}(s_{43})$
is$1:1$. This can be seen by observing that the kinematic factors
$s_{46}$ and $s_{43}$ always show up together through the combination
$s_{46}+s_{43}$ in all of the translation coefficients $\mathcal{C}$
in the $t=3$ sector (see equations from (\ref{eq:transltn-c1}) to
(\ref{eq:transltn-c6}) as well as (\ref{eq:t3})). The only part
of the argument that requires explicit calculation is the ratio between
$J_{3}(s_{46})$ and $J_{4}(s_{36})$. For the purpose of discussion
let us tentatively call them respectively as $X$ and $Y$. From equation
(\ref{eq:t3}) and the definition of $E\left[45,3\right]$ and $E\left[54,3\right]$,
the contribution associated with $J_{3}(s_{46})$ reads
\bea
X & = & \frac{1}{s_{42}s_{52}(s_{45}+s_{42}+s_{52})}\left[s_{52}(s_{45}+s_{42}+s_{52})\left(-T_{3}(245)+T_{3}(425)
\right)\right.\nn
 &  & +s_{52}(s_{52}+s_{56}+s_{53}+s_{54})\left(T_{3}(245)-T_{3}(425)-T_{3}(524)+T_{3}(542)\right)\nonumber \\
 &  & \left.+s_{42}(s_{52}+s_{56}+s_{53}+s_{54})\left(T_{3}(254)-T_{3}(425)+T_{3}(452)-T_{3}(524)\right)\right]
 \nonumber \\
 & = & \left[s_{36}s_{56}A\left(4,2,3,5,6,7\right)\right.
 +s_{36}\left(s_{35}+s_{56}\right)A\left(4,2,5,3,6,7\right)\nonumber \\
 &  & -s_{56}\left(s_{25}+s_{26}+s_{56}+s_{35}+s_{45}\right)A\left(4,3,2,5,6,7\right) +s_{26}\left(s_{35}+s_{45}\right)A\left(4,3,5,2,6,7\right)\nonumber \\
 &  & -s_{36}s_{45}A\left(4,5,2,3,6,7\right) \left.+s_{26}s_{45}A\left(4,5,3,2,6,7\right)\right]\label{eq:j3-1}
\eea

and similarly,
\bea
Y & = & \frac{1}{s_{32}s_{52}(s_{52}+s_{32}+s_{35})}\left[s_{52}(s_{52}+s_{32}+s_{35})\left(-T_{4}(235)
+T_{4}(325)\right)\right.\nn
 &  & +s_{32}(s_{52}+s_{53}+s_{54}+s_{56})\left(T_{4}(253)-T_{4}(325)+T_{4}(352)-T_{4}(523)\right)\nonumber \\
 &  & \left.+s_{52}(s_{52}+s_{53}+s_{54}+s_{56})\left(T_{4}(235)-T_{4}(325)-T_{4}(523)+T_{4}(532)\right)\right]\nonumber \\
 & = & \left[s_{36}s_{56}A\left(4,2,3,5,6,7\right)\right.
 +s_{36}\left(s_{35}+s_{56}\right)A\left(4,2,5,3,6,7\right)\nonumber \\
 &  & -s_{56}\left(s_{25}+s_{26}+s_{56}+s_{35}+s_{45}\right)A\left(4,3,2,5,6,7\right)
 +s_{26}\left(s_{35}+s_{45}\right)A\left(4,3,5,2,6,7\right)\nonumber \\
 &  & -s_{36}s_{45}A\left(4,5,2,3,6,7\right) \left.+s_{26}s_{45}A\left(4,5,3,2,6,7\right)\right]\label{eq:j4-1}
\eea

Now that we have the explicit formulas of the $J_{3}(s_{46})$ and
$J_{4}(s_{36})$ term contributions, it is evident from (\ref{eq:j3-1})
and (\ref{eq:j4-1}) that they are related by an exchange of legs
$3$ and $4$, $Y=\left.X\right|_{3\leftrightarrow4}$. Therefore
to prove $X=-Y$ it suffices to show that $Y$ is antisymmetric with
respect to indices $3$ and $4$. This antisymmetric structure will
become manifest after some nontrivial manipulations, which we perform
in the following.

First of all note that BCJ relation allows us to write
\bea  
 &  & s_{36}s_{56}A\left(4,2,3,5,6,7\right)+s_{36}\left(s_{35}+s_{56}\right)A\left(4,2,5,3,6,7\right)\nn
 &  & =-s_{36}\left(s_{52}+s_{53}+s_{56}\right)A\left(4,5,2,3,6,7\right)-s_{36}\left(s_{52}+s_{53}+s_{56}
 +s_{54}\right)A\left(4,2,3,6,7,5\right)
\eea 
and
\bea 
 &  & s_{26}\left(s_{35}+s_{45}\right)A\left(4,3,5,2,6,7\right)+s_{26}s_{45}A\left(4,5,3,2,6,7\right)\nn
 &  & =-s_{26}\left(s_{35}+s_{45}+s_{25}\right)A\left(4,3,2,5,6,7\right)-s_{26}\left(s_{35}+s_{45}+s_{25}+s_{65}\right)A\left(4,3,2,6,5,7\right).
\eea 
Plugging the above two identities into the expression for $Y$, we
have
\bea
Y & = & -s_{36}\left(s_{52}+s_{53}+s_{56}+s_{54}\right)\left[A\left(4,5,2,3,6,7\right)+A\left(4,2,3,6,7,5\right)
\right]\nn
 &  & -\left[s_{26}\left(s_{35}+s_{45}+s_{25}\right)+s_{56}\left(s_{25}+s_{26}+s_{56}+s_{35}+s_{45}\right)\right]
 A\left(4,3,2,5,6,7\right)\nn
 &  & -s_{26}\left(s_{35}+s_{45}+s_{25}+s_{65}\right)A\left(4,3,2,6,5,7\right)\nn
 & = & -\left(s_{52}+s_{53}+s_{56}+s_{54}\right)\nn
 &  & \times\left[s_{36}A\left(4,5,2,3,6,7\right)+s_{36}A\left(4,2,3,6,7,5\right)+\left(s_{26}
 +s_{56}\right)A\left(4,3,2,5,6,7\right)+s_{26}A\left(4,3,2,6,5,7\right)\right]\nn
 & = & s_{57}\left[s_{36}A\left(4,5,2,3,6,7\right)+s_{36}A\left(4,2,3,6,7,5\right)+\left(s_{26}
 +s_{56}\right)A\left(4,3,2,5,6,7\right)+s_{26}A\left(4,3,2,6,5,7\right)\right]\nn
\eea
Further using BCJ relation identifies the sum of last two terms above
with
\bea 
 &  & \left(s_{26}+s_{56}\right)A\left(4,3,2,5,6,7\right)+s_{26}A\left(4,3,2,6,5,7\right)\nn
 &  & =-\left(s_{26}+s_{56}+s_{76}\right)A\left(4,3,2,5,7,6\right)-\left(s_{26}+s_{56}+s_{76}+s_{46}
 \right)A\left(4,6,3,2,5,7\right)\nn
 &  & =\left(s_{36}+s_{46}\right)A\left(4,3,2,5,7,6\right)+s_{36}A\left(4,6,3,2,5,7\right)
\eea  
Therefore Y simplifies as
\begin{eqnarray}
Y & = & s_{57}s_{36}\left[A\left(4,5,2,3,6,7\right)+A\left(4,2,3,6,7,5\right)+A\left(4,3,2,5,7,6\right)+A\left(4,6,3,2,5,7\right)\right]\nonumber \\
 &  & +s_{57}s_{46}A\left(4,3,2,5,7,6\right)\nonumber \\
 & = & s_{57}\left[s_{46}A\left(4,3,2,5,7,6\right)-s_{36}A\left(3,4,2,5,7,6\right)\right]\label{eq:formula-y}
\end{eqnarray}
where we used $U(1)$ decoupling identity to substitute the summation
in the first line with a single amplitude. The final simplified formula
of $Y$ is manifestly antisymmetric under the exchange of indices
$3$ and $4$, and we conclude that $X+Y=0$ as claimed.



\begin{thebibliography}{References}


\bibitem{Low}
F.E. Low, Phys.\ Rev.\  {\bf 96}, 1428 (1954), Phys.\ Rev.\  {\bf 110}, 974 (1958);
M. Gell-Mann and M.L. Goldberger, Phys.\ Rev.\  {\bf 96}, 1433 (1954); S. Saito, Phys.\ Rev.\  {\bf 184}, 1894 (1969);


\bibitem{Weinberg:1964ew}
  S.~Weinberg,
  ``Photons and Gravitons in s Matrix Theory: Derivation of Charge Conservation and Equality of Gravitational and Inertial Mass,''
  Phys.\ Rev.\  {\bf 135} (1964) B1049.

  S.~Weinberg,
  ``Infrared photons and gravitons,''
  Phys.\ Rev.\  {\bf 140} (1965) B516.

\bibitem{Strominger:2013jfa}
  A.~Strominger,
  ``On BMS Invariance of Gravitational Scattering,''
  arXiv:1312.2229 [hep-th].

\bibitem{He:2014laa}
  T.~He, V.~Lysov, P.~Mitra and A.~Strominger,
  ``BMS supertranslations and Weinberg's soft graviton theorem,''
  arXiv:1401.7026 [hep-th].

\bibitem{Kapec:2014opa}
  D.~Kapec, V.~Lysov, S.~Pasterski and A.~Strominger,
  ``Semiclassical Virasoro Symmetry of the Quantum Gravity S-Matrix,''
  arXiv:1406.3312 [hep-th].

\bibitem{Britto:2004ap}
  R.~Britto, F.~Cachazo and B.~Feng,
  ``New recursion relations for tree amplitudes of gluons,''
  Nucl.\ Phys.\ B {\bf 715} (2005) 499
  [hep-th/0412308].

\bibitem{Britto:2005fq}
  R.~Britto, F.~Cachazo, B.~Feng and E.~Witten,
  ``Direct proof of tree-level recursion relation in Yang-Mills theory,''
  Phys.\ Rev.\ Lett.\  {\bf 94} (2005) 181602
  [hep-th/0501052].


\bibitem{Cachazo:2014fwa}
  F.~Cachazo and A.~Strominger,
  ``Evidence for a New Soft Graviton Theorem,''
  arXiv:1404.4091 [hep-th].


\bibitem{Bern:1998sv}
  Z.~Bern, L.~J.~Dixon, M.~Perelstein and J.~S.~Rozowsky,
  Nucl.\ Phys.\ B {\bf 546}, 423 (1999)
  [hep-th/9811140].



\bibitem{White}
  E.~Laenen, G.~Stavenga and C.~D.~White,
  JHEP {\bf 0903}, 054 (2009)
  [arXiv:0811.2067 [hep-ph]].

  E.~Laenen, L.~Magnea, G.~Stavenga and C.~D.~White,
  JHEP {\bf 1101}, 141 (2011)
  [arXiv:1010.1860 [hep-ph]].

  C.~D.~White,
  JHEP {\bf 1105}, 060 (2011)
  [arXiv:1103.2981 [hep-th]].


\bibitem{Casali:2014xpa}
  E.~Casali,
  ``Soft sub-leading divergences in Yang-Mills amplitudes,''
  arXiv:1404.5551 [hep-th].


\bibitem{Broedel:2014fsa}
  J.~Broedel, M.~de Leeuw, J.~Plefka and M.~Rosso,
  ``Constraining subleading soft gluon and graviton theorems,''
  arXiv:1406.6574 [hep-th].

\bibitem{Bern:2014vva}
  Z.~Bern, S.~Davies, P.~Di Vecchia and J.~Nohle,
  ``Low-Energy Behavior of Gluons and Gravitons from Gauge Invariance,''
  arXiv:1406.6987 [hep-th].

\bibitem{White:2014qia}
  C.~D.~White,
  ``Diagrammatic insights into next-to-soft corrections,''
  arXiv:1406.7184 [hep-th].

\bibitem{Larkoski:2014hta}
  A.~J.~Larkoski,
  ``Conformal Invariance of the Subleading Soft Theorem in Gauge Theory,''
  arXiv:1405.2346 [hep-th].



\bibitem{Schwab:2014xua}
  B.~U.~W.~Schwab and A.~Volovich,
  ``Subleading soft theorem in arbitrary dimension from scattering equations,''
  arXiv:1404.7749 [hep-th].

\bibitem{Afkhami-Jeddi:2014fia}
  N.~Afkhami-Jeddi,
  ``Soft Graviton Theorem in Arbitrary Dimensions,''
  arXiv:1405.3533 [hep-th].

\bibitem{Zlotnikov:2014sva}
  M.~Zlotnikov,
  ``Sub-sub-leading soft-graviton theorem in arbitrary dimension,''
  arXiv:1407.5936 [hep-th].

\bibitem{Kalousios:2014uva}
  C.~Kalousios and F.~Rojas,
  ``Next to subleading soft-graviton theorem in arbitrary dimensions,''
  arXiv:1407.5982 [hep-th].

\bibitem{Bern:2014oka}
  Z.~Bern, S.~Davies and J.~Nohle,
  ``On Loop Corrections to Subleading Soft Behavior of Gluons and Gravitons,''
  arXiv:1405.1015 [hep-th].

\bibitem{He:2014bga}
  S.~He, Y.~-t.~Huang and C.~Wen,
  ``Loop Corrections to Soft Theorems in Gauge Theories and Gravity,''
  arXiv:1405.1410 [hep-th].

\bibitem{Cachazo:2014dia}
  F.~Cachazo and E.~Y.~Yuan,
  ``Are Soft Theorems Renormalized?,''
  arXiv:1405.3413 [hep-th].


\bibitem{Schwab:2014fia}
  B.~U.~W.~Schwab,
  ``Subleading Soft Factor for String Disk Amplitudes,''
  arXiv:1406.4172 [hep-th].

\bibitem{Bianchi:2014gla}
  M.~Bianchi, S.~He, Y.~-t.~Huang and C.~Wen,
  ``More on Soft Theorems: Trees, Loops and Strings,''
  arXiv:1406.5155 [hep-th].


\bibitem{Adamo:2014yya}
  T.~Adamo, E.~Casali and D.~Skinner,
  arXiv:1405.5122 [hep-th].

\bibitem{Geyer:2014lca}
  Y.~Geyer, A.~E.~Lipstein and L.~Mason,
  arXiv:1406.1462 [hep-th].

\bibitem{Bern:2008qj}
  Z.~Bern, J.~J.~M.~Carrasco and H.~Johansson,
  Phys.\ Rev.\ D {\bf 78}, 085011 (2008)
  [arXiv:0805.3993 [hep-ph]].


\bibitem{Cachazo:2013hca}
  F.~Cachazo, S.~He and E.~Y.~Yuan,
  arXiv:1307.2199 [hep-th].


\bibitem{Cachazo:2013iea}
  F.~Cachazo, S.~He and E.~Y.~Yuan,
  JHEP {\bf 1407}, 033 (2014)
  [arXiv:1309.0885 [hep-th]].


 \bibitem{KLT} H. Kawai, D. Lewellen and H. Tye, "A Relation Betwwen Tree
Amplitudes of Closed and Open Strings", Nucl.Phys.B269 (1986)1.

\bibitem{BjerrumBohr:2010ta}
  N.~E.~J.~Bjerrum-Bohr, P.~H.~Damgaard, B.~Feng and T.~Sondergaard,
  ``Gravity and Yang-Mills Amplitude Relations,''
  Phys.\ Rev.\ D {\bf 82} (2010) 107702
  [arXiv:1005.4367 [hep-th]].

\bibitem{BjerrumBohr:2010yc}
  N.~E.~J.~Bjerrum-Bohr, P.~H.~Damgaard, B.~Feng and T.~Sondergaard,
  ``Proof of Gravity and Yang-Mills Amplitude Relations,''
  JHEP {\bf 1009} (2010) 067
  [arXiv:1007.3111 [hep-th]].

\bibitem{BjerrumBohr:2010hn}
  N.~E.~J.~Bjerrum-Bohr, P.~H.~Damgaard, T.~Sondergaard and P.~Vanhove,
  ``The Momentum Kernel of Gauge and Gravity Theories,''  JHEP {\bf 1101} (2011) 001  [arXiv:1010.3933 [hep-th]].  





\bibitem{Cachazo:2012da}
  F.~Cachazo and Y.~Geyer,
  arXiv:1206.6511 [hep-th].

\bibitem{Cachazo:2012kg}
  F.~Cachazo and D.~Skinner,
  Phys.\ Rev.\ Lett.\  {\bf 110}, no. 16, 161301 (2013)
  [arXiv:1207.0741 [hep-th]].

\bibitem{Cachazo:2013zc}
  F.~Cachazo,
  arXiv:1301.3970 [hep-th].







































\end{thebibliography}
\end{document}